\documentclass[12pt,preprint,a4paper]{aastex}
\usepackage{times}
\usepackage{graphics,epsf}
\usepackage{amsmath}                % American Mathematical Society package
\usepackage{amsfonts}               % American Mathematical Society fonts
\usepackage{amssymb}                % American Mathematical Society symbol
\usepackage{epsfig}                 % EPS figures
\usepackage{graphicx}
\usepackage{url}
\usepackage{tabularx}
\usepackage{booktabs}
\usepackage{epstopdf}

\def \cm{~\rm{cm}}
\def \s{~\rm{s}}
\def \km{~\rm{km}}

\def \g{~\rm{g}}

\def \erg{~\rm{erg}}

\def \days{~\rm{day}}

\begin{document}

\title{An outburst powered by the merging of two stars inside the envelope of a giant}

\author{Shlomi Hillel\altaffilmark{1}, Ron Schreier\altaffilmark{1}, and  Noam Soker\altaffilmark{1}}
\altaffiltext{1}{Department of Physics, Technion -- Israel
Institute of Technology, Haifa 32000, Israel; shlomihi@tx.technion.ac.il, ronsr@physics.technion.ac.il, soker@physics.technion.ac.il}

\begin{abstract}
We conduct three-dimensional hydrodynamical simulations of energy deposition into the envelope
of a red giant star as a result of the merger of two close main sequence stars or brown dwarfs,
and show that the outcome is a highly non-spherical outflow.
Such a violent interaction of a triple stellar system can explain the formation of `messy', i.e., lacking any kind of symmetry, planetary nebulae (PNe) and similar nebulae around evolved stars.
We do not simulate the merging process, but simply assume that after the tight binary system enters the envelope of the giant star the interaction with the envelope causes the two components, stars or brown dwarfs, to merge and liberate gravitational energy.
We deposit the energy over a time period of about nine hours, which is about one per cent of the orbital period of the merger product around the centre of the giant star. The ejection of the fast hot gas and its collision with previously ejected mass are very likely to lead to a transient event,
i.e., an intermediate luminosity optical transient (ILOT).
\end{abstract}


% ==========================================================
\section{INTRODUCTION}
\label{sec:intro}
% ==========================================================

The mass loss rate and outflow geometry from a giant star can be substantially influenced by the presence of close objects, i.e., planets, brown dwarfs and/or stars, that interact with the giant star.
Planetary nebulae (PNe) is the most studied group of objects that are shaped by the interaction of their progenitor, an asymptotic giant branch (AGB) star,
with stellar companions (e.g, \citealt{BondLivio1990, Bond2000, DeMarco2015, Zijlstra2015, JonesBoffin2017b}) or with planets (e.g, \citealt{DeMarcoSoker2011}).
Following early theoretical and observational studies \citep{LivioShaviv1975, Bondetal1978, FabianHansen1979, Morris1981, Paczynski1985,IbenTutukov1989, BondLivio1990, Hanetal1995}
observations of tens of PNe with central binary systems, in particular in recent years
(e.g., \citealt{Akrasetal2015, Alleretal2015a, Alleretal2015b, Boffin2015, Corradietal2015, Decinetal2015, DeMarcoetal2015,
Douchinetal2015, Fangetal2015, Gorlovaetal2015, Hillwigetal2015, Jones2015, Jonesetal2015, Manicketal2015,
Martinezetal2015, Miszalskietal2015, Mocniketal2015, Montezetal2015, Jonesetal2016, Chiotellisetal2016,
Akrasetal2016, GarciaRojasetal2016, Jones2016, Hillwigetal2016a,
Bondetal2016, Chenetal2016, Madappattetal2016, Alietal2016, Hillwigetal2016b, Jonesetal2017}, for a partial sample just from 2015 on), have solidified the binary interaction model for shaping PNe. Single stars cannot lead to the variety of shapes of PNe (e.g., \citealt{SokerHarpaz1992, NordhausBlackman2006, GarciaSeguraetal2014}). But what about triple stellar (or sub-stellar) systems? Triple systems attract attention in other types of processes and objects (e.g., \citealt{EggletonVerbunt1986, MichaelyPerets2014}).

A small number of papers discussed the shaping of PNe by triple stellar systems (e.g., \citealt{Sokeretal1992, Soker1994,
Bondetal2002, Exteretal2010, Danehkaretal2013, Soker2016triple, BearSoker2017}). Note though that the claim made by \cite{Exteretal2010} for a triple stellar system inside a PN was refuted recently by \cite{JonesBoffin2017}.
\cite{Exteretal2010} proposed that the progenitor of the PN SuWt 2 engulfed a tight binary system of two A-type stars, and that the A-stars binary system survived the common envelope evolution. They further suggested that triple stellar interaction might eject a high-density equatorial ring (see also \citealt{Bondetal2002}). However, \cite{JonesBoffin2017} concluded recently that the binary system of A-type stars is a field star system, which happens to be along the line of sight to SuWt 2.

We consider triple systems for which two orbital planes can be defined.
One is that of the more tight binary system, and the second one is that of the tight binary system motion around the centre of mass with the third object.
When the two orbital planes are inclined to each other, the mass loss geometry is likely to depart from any kind of symmetry, and so is the descendant PN. The PN will lack any kind of symmetry;
nor point-symmetry, nor axial-symmetry, and nor mirror symmetry.
We term this a `messy PN' \citep{Soker2016triple, BearSoker2017}.

Most nebulae do not require triple stellar interaction. \cite{Portegies2016} suggest that the Great Eruption of Eta Carinae was caused by a merger in a triple stellar system.
But there are claims that a binary system alone can explain the the Great Eruption of Eta Carinae and the formation of the bipolar structure of the Homunculus that was formed during the Great Eruption \citep{KashiSoker2010}. The binary model can explain also earlier \citep{Kiminkietal2016} and later (the Lesser Eruption; \citealt{Humphreysetal1999}) outbursts.
In the present study we consider a different type of triple-stellar evolution than that considered by \cite{Portegies2016}.

When the two orbital planes of the triple system coincide the interaction leads to a mass loss process that has a symmetry plane,
although it might depart from axisymmetry.
Binary systems can also lead to departure from axisymmetry (e.g., \citealt{SokerHadar2002}). However, binary systems are unlikely to form messy PNe that lack any symmetry.
In a recent paper \cite{Chenetal2017} conduct 3D hydrodynamical simulation of mass transfer from an AGB star to a close companion. They demonstrate that when the AGB star losses mass in a short (5 years) burst, a very complicated and highly asymmetric mass loss takes place. This binary process still retains a symmetry about the equatorial plane. They do not study the launching of jets. Precessing jets could lead to a departure from any symmetry in that case of a short burst, and lead to the formation of a somewhat messy PN.

In many cases the more compact companion accretes mass from the AGB star.
The accreted mass has high specific angular momentum and it forms an accretion disk around the companion,
that in turn
launches jets that shape the descendant PN
(e.g., \citealt{Morris1987, Soker1990AJ, SahaiTrauger1998, AkashiSoker2008,
Boffinetal2012, HuarteEspinosaetal2012, Balicketal2013, Miszalskietal2013,
Tocknelletal2014, Huangetal2016, Sahaietal2016, RechyGarciaetal2017},
out of many more papers).
Triple stellar systems can also launch jets.
\cite{AkashiSoker2017} simulate the flow structure described by \cite{Soker2004}, where a tight binary system orbits an AGB star and accretes mass from the AGB wind. Because the orbital plane of the tight binary system is not parallel to the orbital plane around the AGB star in that setting, the jets' axis is not perpendicular to the orbital plane. This triple stellar interaction forms a messy nebula \citep{AkashiSoker2017}.

We here study another process that is likely to form a messy nebula.
This is the evolutionary channel of a merger of a tight binary system inside the envelope of a giant star \citep{Soker2016triple}.
The components of the tight binary system can be stars, brown dwarfs, or massive planets.
Although the tight binary system influences the mass loss process before it enters the envelope, e.g.,
by tidal interaction, spinning-up the envelope, and launching jets,
such a full study requires huge amount of computer resources.
For resources constraints, we simply take the energy that is expected to be released by the merger process,
and deposit it inside the envelope of the giant star.
In a future paper we will also consider some of the other effects mentioned above.
We describe the initial setting and the three-dimensional numerical code in section \ref{sec:numerical}.
We then describe the numerical results in section \ref{sec:results}.
We summarize our results in section \ref{sec:summary}.

% ==========================================================
\section{NUMERICAL SET-UP}
 \label{sec:numerical}
% ==========================================================

We run the stellar evolution code
\texttt{MESA} \citep{Paxtonetal2011, Paxtonetal2013, Paxtonetal2015}
to obtain a spherical AGB model with zero-age-main-sequence mass of $M_{ZAMS}=4 M_\sun$.
We let the star evolve until it reaches the AGB stage after $3\times10^8$ years.
At that time the  stellar radius is $R_{AGB}=100\,R_{\sun}$, and its effective temperature is $T_{eff}= 3400K$.
We then import the spherical AGB model, namely, the profiles of density and pressure into the three-dimensional hydrodynamical code {\sc pluto} \citep{Mignone2007}.
The full grid is taken as a cube with side length of $400\,R_{\sun}$.
We employ an adaptive-mesh-refinement (AMR) grid with five refinement levels.
The base grid resolution is $1/64$ of the grid length
(i.e. $4.35\times 10^{11} \cm$),
and the highest resolution is $2^4$ times smaller (i.e. $2.72\times 10^{10} \cm$). The centre of the AGB star and the centre of the grid coincide at $(x,y,z)=(0,0,0)$.
We use an equation of state of an ideal gas with adiabatic index $\gamma=5/3$.

The code runs on time steps determined by the Courant - Friedrichs - Lewy (CFL) condition, and thus the high densities in the centre of the primary requires very short time steps.
As we are not interested in the inner parts of the AGB star,
its inner $5\%$ in radius ($5\,R_{\sun}$) were replaced with a constant density, pressure, and temperature sphere. This allows the code to run with larger time steps than those required for simulating this inner region.
The dynamic field of the local gravitational acceleration were inserted as if the inner parts of the AGB star were not truncated. The initial stellar structure is in hydrostatic equilibrium. We keep the gravitational field at its initial value (at $t=0$) throughout the entire calculation. Namely, we do not include neither the change in gravity that results from the deformed envelope nor the gravity of the merger product.

We consider a scenario where a tight binary system composed of two main sequence stars or two massive brown dwarfs of masses $M_2$ and $M_3$, enters the envelope of the AGB star, i.e., enters a common envelope evolution (CEE). Following \cite{Soker2016triple},  we assume that during the CEE gravitational drag and mass accretion cause the tight binary system both to spiral-in toward the AGB centre, and might cause the two main sequence stars (or brown dwarfs) to merge with each other.
The simulation starts with the assumed merger of the two stars
of the tight binary system at $r=70\, R_\sun$. We set the coordinate system such that at $t=0$ the merger takes place at
$(x,y,z)_{\rm mer-0}=(70 R_\odot,0,0)$.
At that orbital separation the Keplerian orbital period
of the merger product around the AGB core is about $36 \days$.

A few words here are in place on the initial conditions. We start the simulations with the merger process taking place inside the envelope, and ignore the interaction before that stage, e.g., the spin-up of the envelope by the tight binary system, jets that might have been launched by an accretion disk around one (or two) star of the tight binary system. In addition, the entrance to the common envelope can substantially distort the envelope and the mass loss process (e.g., \citealt{MorrisPodsiadlowski2009} for a scenario explaining the rings of SN 1987A). The reason for this initial set up are numerical limitations. However, the inclusion of envelope rotation and envelope expansion due to the spiraling-in process, would most likely make the envelope more vulnerable to perturbations by the merger process. Namely, even a merger of two brown dwarfs or very low mass main sequence stars (see also below) will lead to a very messy PN.

The merger process liberates an energy of $E_{\rm mer} \approx 10^{48} \erg (M_2M_3/M^2_\odot)(R/R_\odot)^{-1}$, where $R$ is the radius of the larger star of the two merging objects.
A large fraction of the energy released in the merger process will be channeled to inflate the merger product, and the rest will go into the envelope of the giant star. We do not know what this fraction is, but as a conservative approach we assume that only a small fraction of the merger energy is deposited in the envelope of the giant star. If this fraction is larger than what we assume, then the same outcome can result from the merger of an even lower mass tight binary system.
In the first case, designated as the fiducial run, we inject a mass of $M_{\rm mer} = 0.1\,M_\odot$ into the AGB envelope, and that mass carries an energy of $E_{\rm mer} = 5 \times 10^{45} \erg$.
This energy corresponds to several percents of the energy that two main sequence stars of masses $0.2 M_\odot$ liberate. This also corresponds to the energy that is liberated from the merger of two massive brown dwarfs, but the brown dwarfs will inject less mass into the envelope.
We simulated a second case with an injected mass of
$M_{\rm mer} = 0.1\,M_\odot$, but with a larger energy of $E_{\rm mer} = 2.5 \times 10^{46} \erg$.

The merger remnant continues its Keplerian motion and exerts gravitational force on the envelope. We do not consider this in the present paper. The gravitational influence of the tight binary system before merger, and of the merger product after merger, will be studied in a future paper.

The energy and mass are injected over a time period of $T_{\rm mer} = 8.8$~hours, and within a sphere of $R_{\rm mer} = 0.2\,R_{\sun}$, with a diameter of 6 basic cells. The energy is inserted as a thermal energy of the injected gas. The injection time is taken to be several times the orbital period of the tight binary system.
During the injection time period, the merger product has moved a distance of about 0.01 times the circumference (1 per cent of an orbit).

To verify numerical stability, we run the simulation without the binary for ten dynamical times of the giant star. We found no noticeable change in the stellar variables,  i.e. $T(r)$, $P(r)$, and $\rho(r)$.

We end the simulations when a relatively significant amount of mass has left the grid, as we cannot follow the fall back of the bound gas.

% ==========================================
\section{RESULTS}
 \label{sec:results}
%===========================================

\begin{figure}[bp]
\centering
\begin{tabular}{cc}
\includegraphics[width=6.5cm]{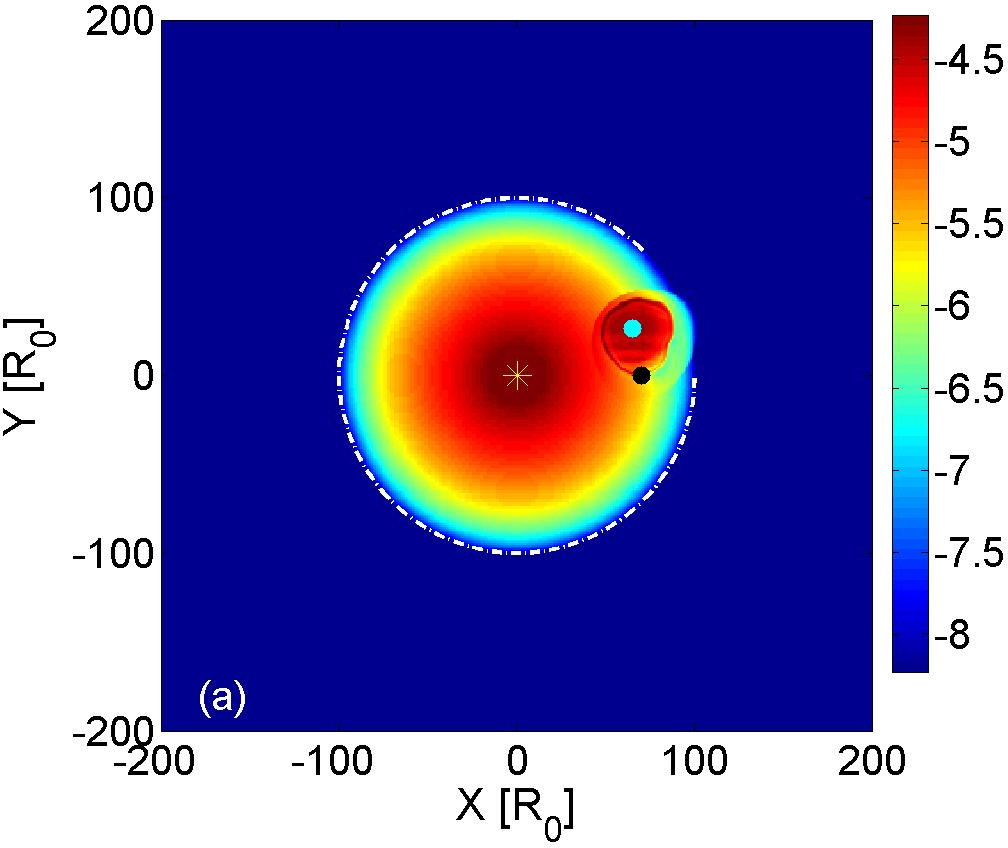}&
\includegraphics[width=6.5cm]{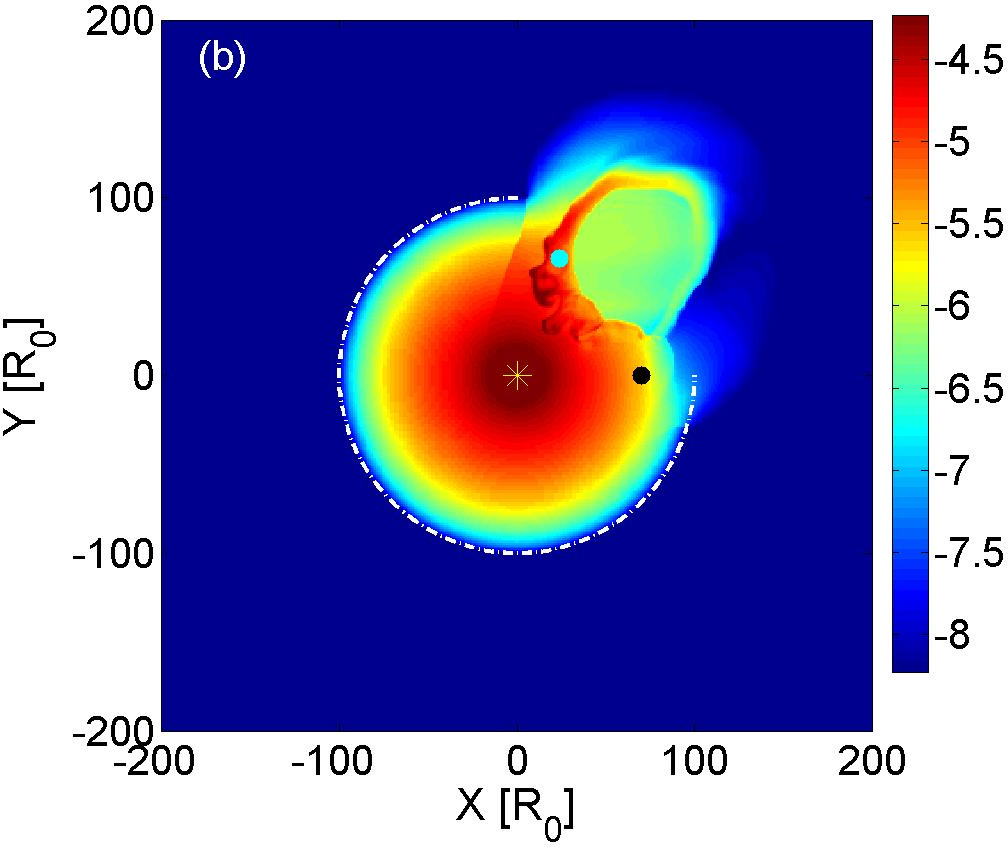}\\
\includegraphics[width=6.5cm]{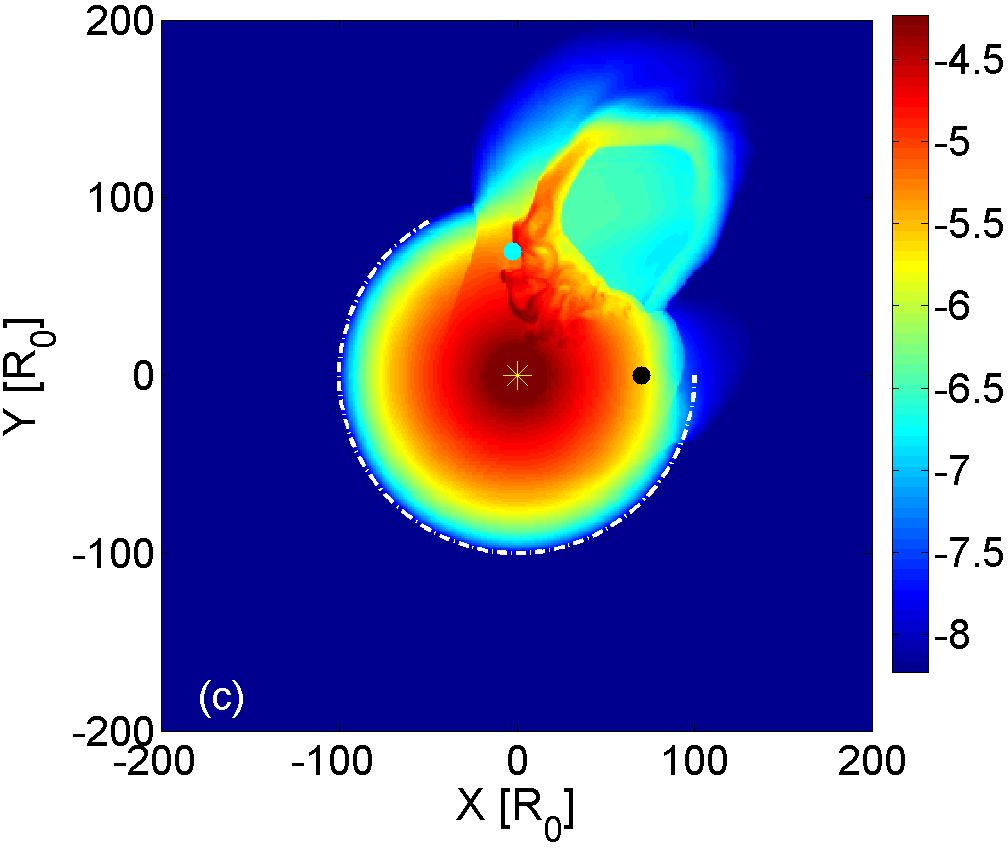}&
\includegraphics[width=6.5cm]{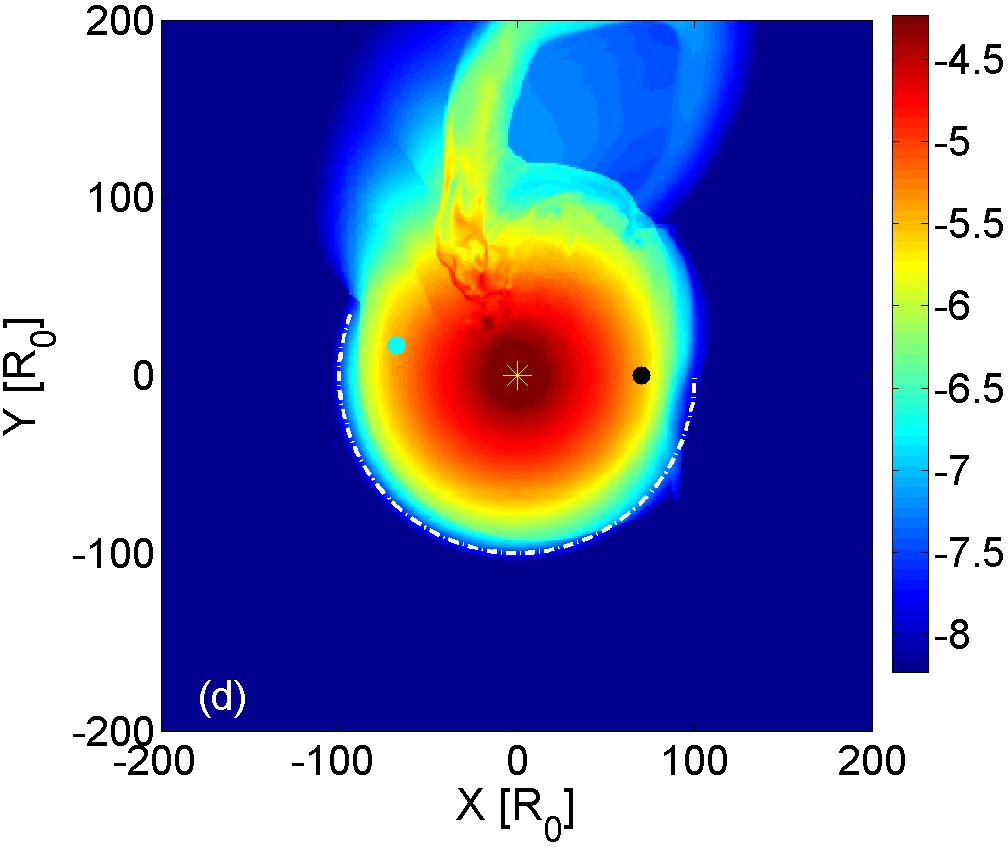}\\
\includegraphics[width=6.5cm]{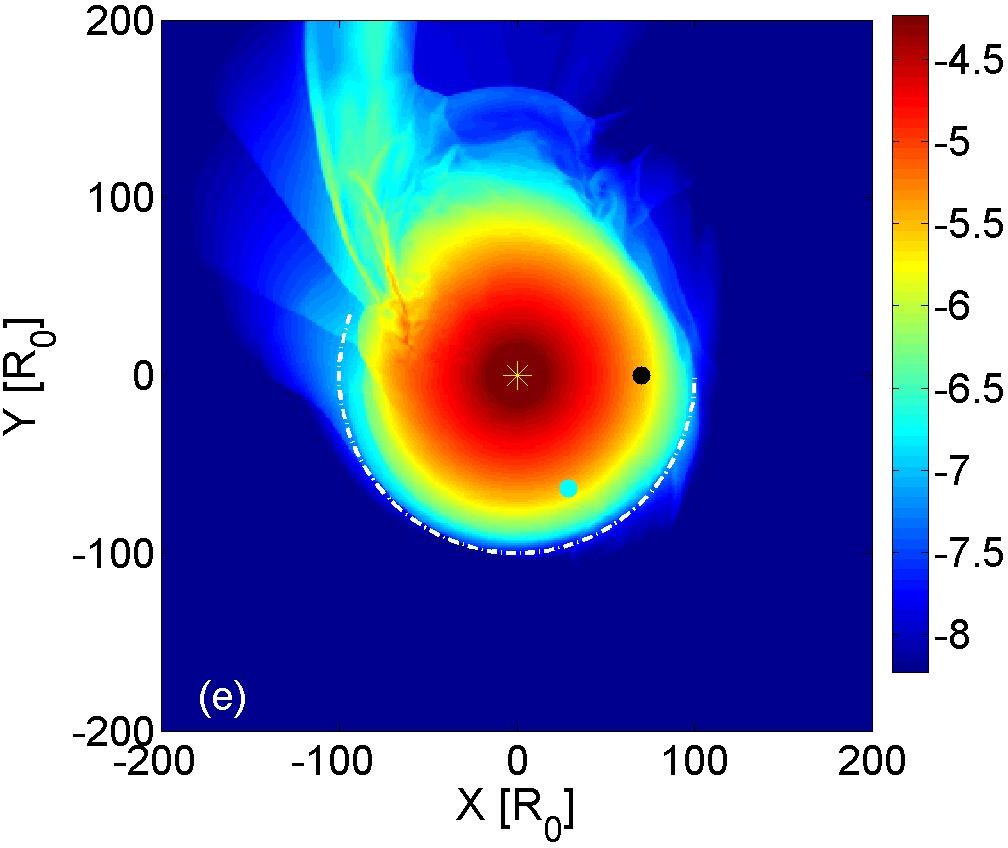}&
\includegraphics[width=6.5cm]{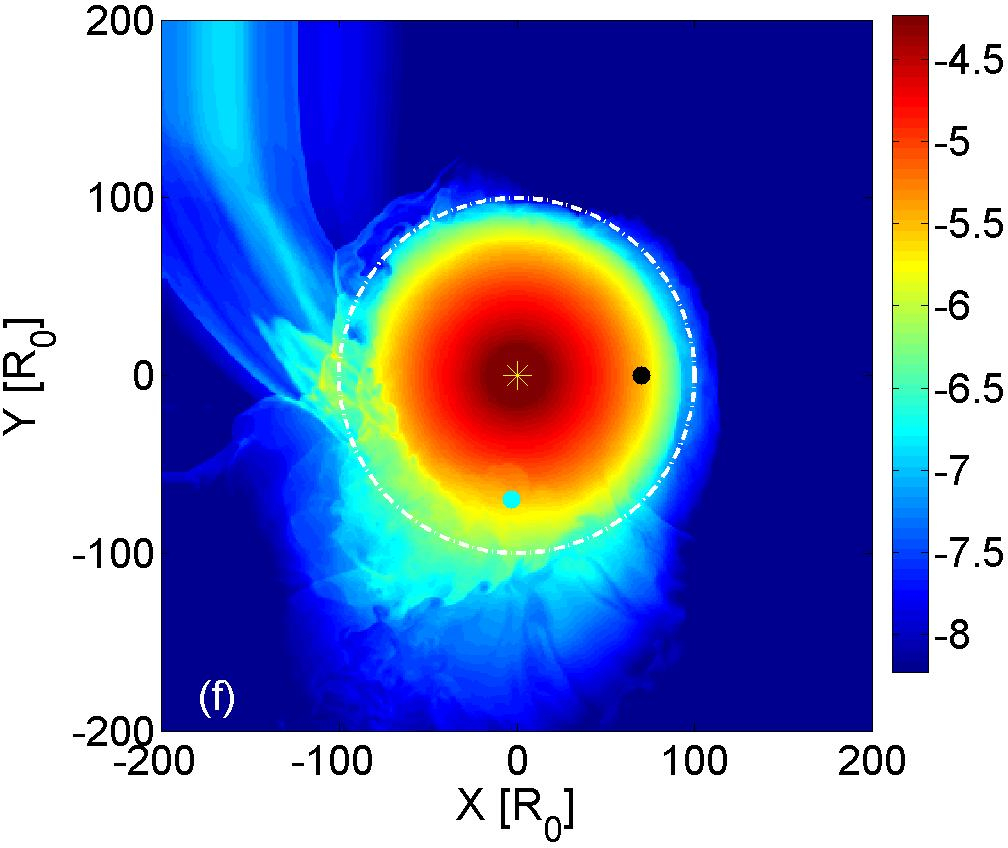}
\end{tabular}
\caption{The evolution of density for the fiducial run with injected energy of $E_{\rm mer} = 5 \times 10^{45} \erg$, in the orbital plane $z=0$ at six times of (a) $t=2$, (b) $7$, (c) $9$, (d) $16$, (e) $29$, and (f) $62$~days, from top to bottom and from left to right. The colour contours are in logarithmic scale as indicated by the colour bars and in units of $\g \cm^{-3}$. A yellow asterisk marks the centre of the AGB star that is also the centre of the grid. The dashed-dotted white line marks the initial surface of the AGB star, the black dot is the location of the merger process, and the cyan circle marks the location of the merger product as it orbits inside the envelope (counterclockwise). Distances on the axes are in $R_{\sun}$.}
  \label{fig:XYslices}
    \end{figure}

\begin{figure}[ht]
\centering
\begin{tabular}{cc}
\includegraphics[width=6.5cm]{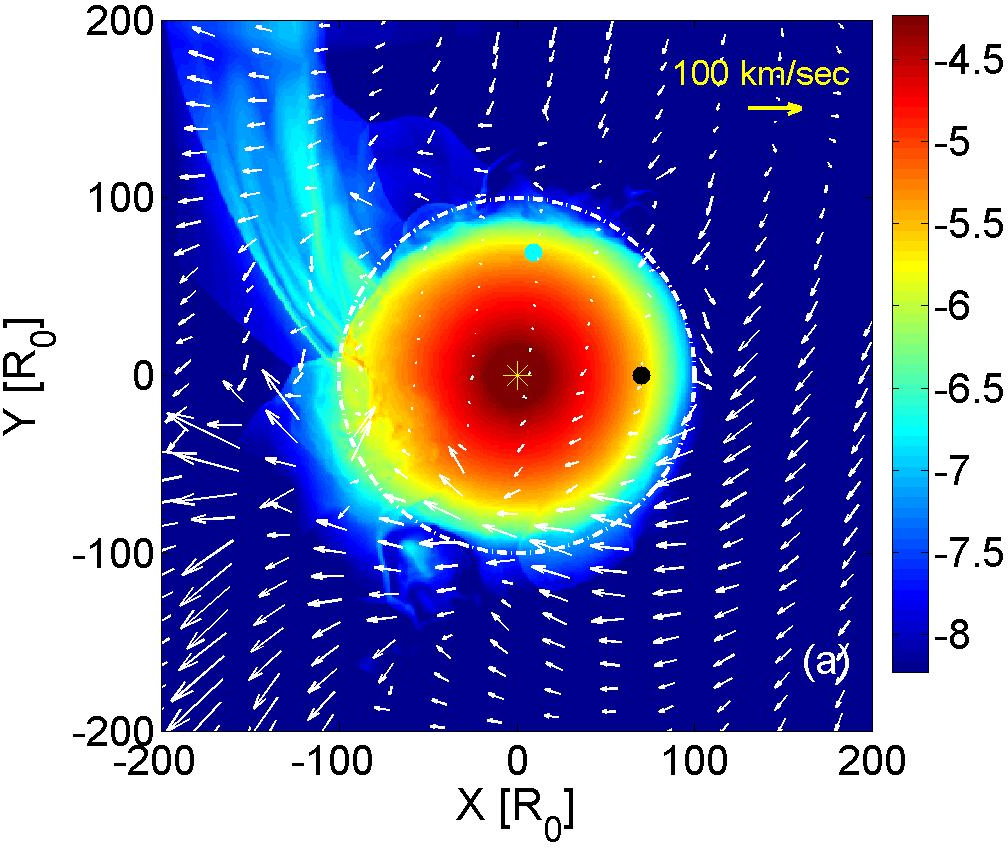}&
\includegraphics[width=6.5cm]{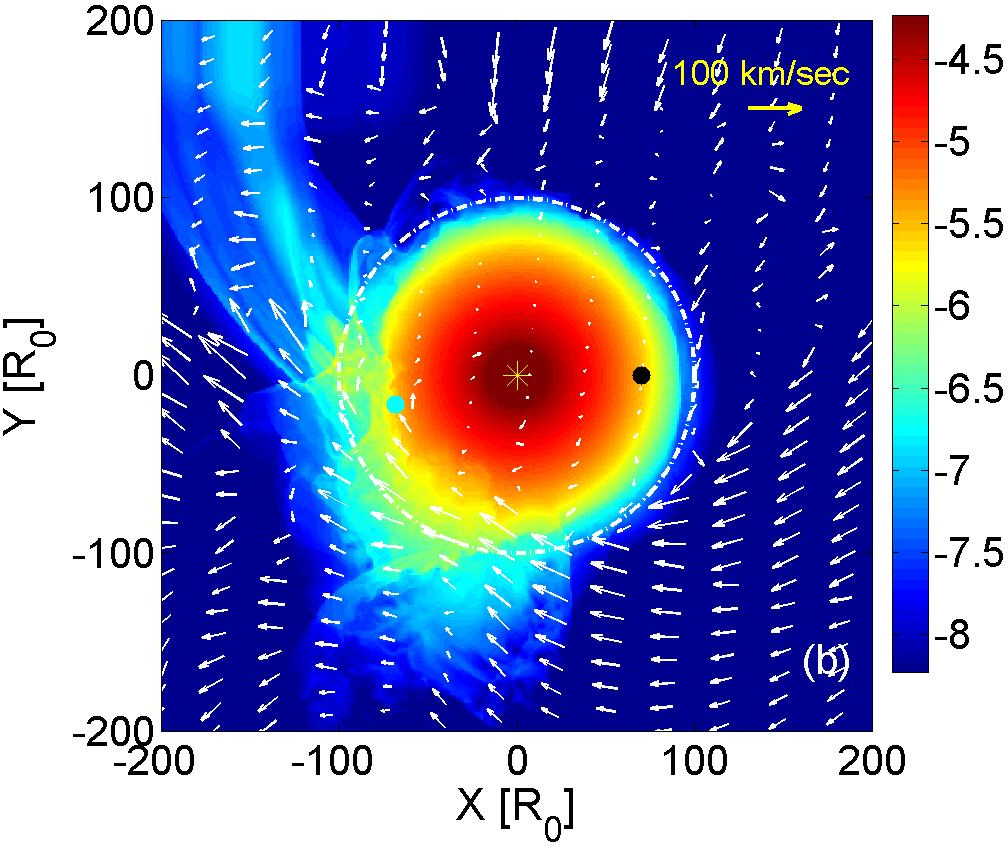}\\
\includegraphics[width=6.5cm]{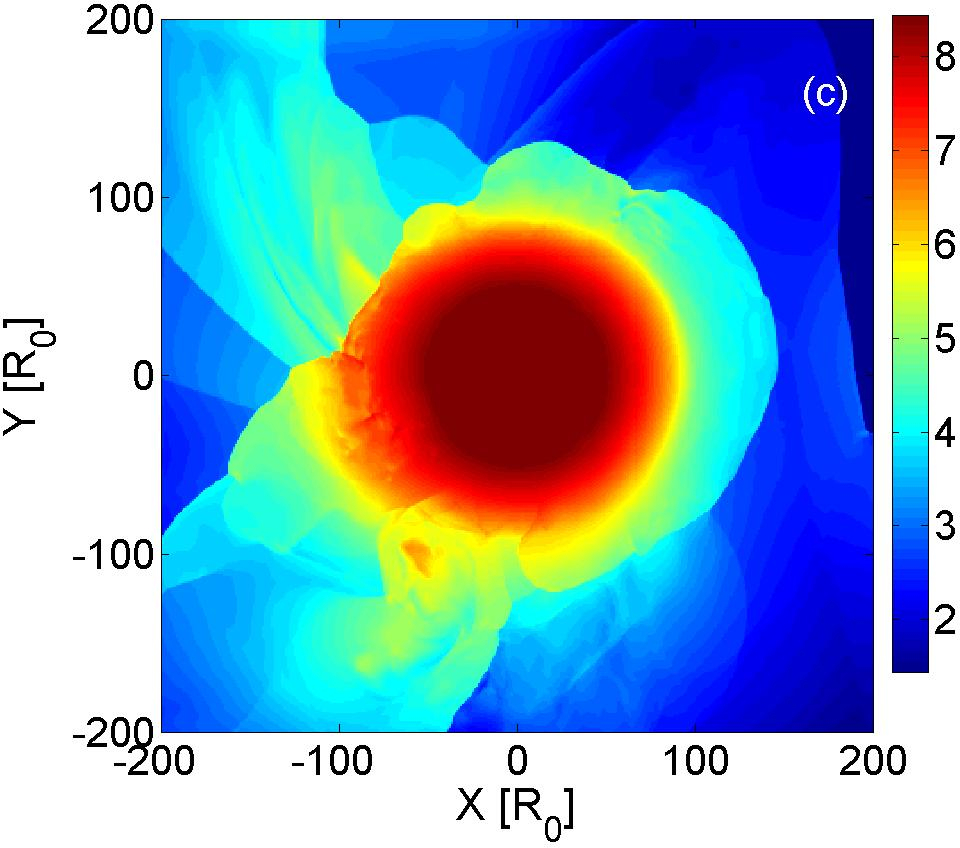}&
\includegraphics[width=6.5cm]{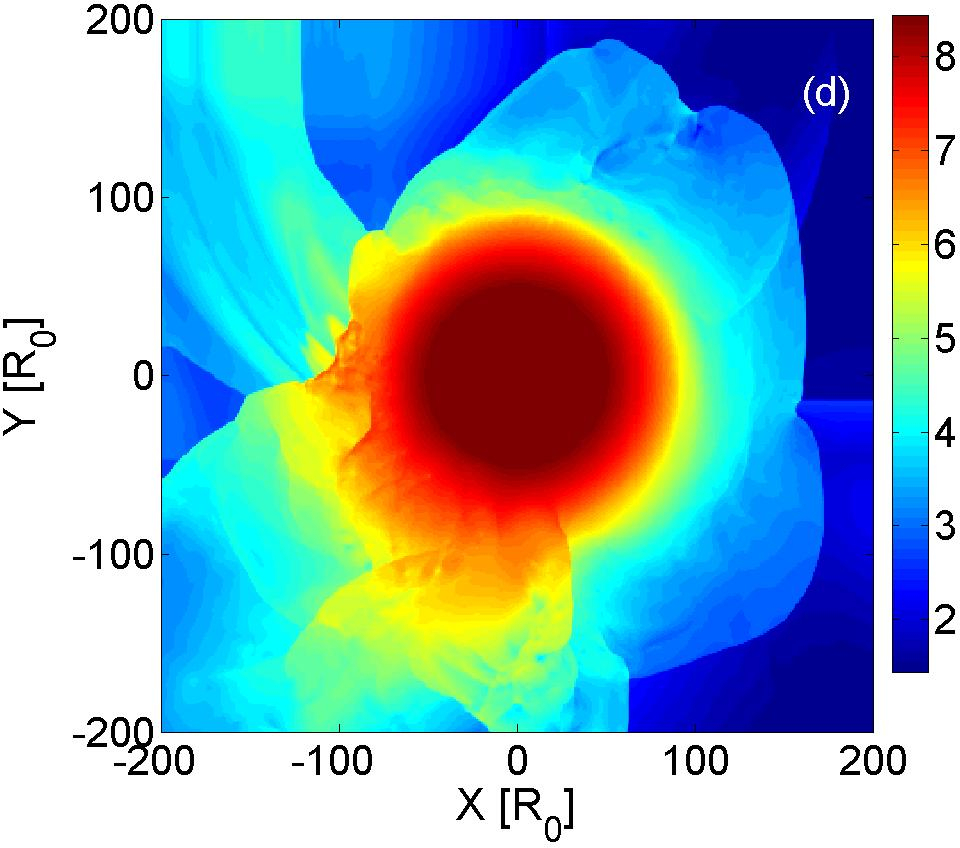}
\end{tabular}
\caption{Density (upper panels) and pressure (bottom panels) maps at $t=44 \days$ (left panels) and at $t=55 \days$ (right panels) in the orbital plane $z=0$. On the density maps we added velocity vectors. The length of each arrow is proportional to the velocity magnitude.  The yellow arrow in the upper right corner represents a velocity of $100 \km \s^{-1}$. The maximum velocities are $208 \km \s^{-1}$ at  $t=44 \days$ and $130 \km \s^{-1}$ at $t=55 \days$.
}
  \label{fig:XYPressure}
    \end{figure}

\begin{figure}[]
\centering
\begin{tabular}{cc}
\includegraphics[width=6.5cm]{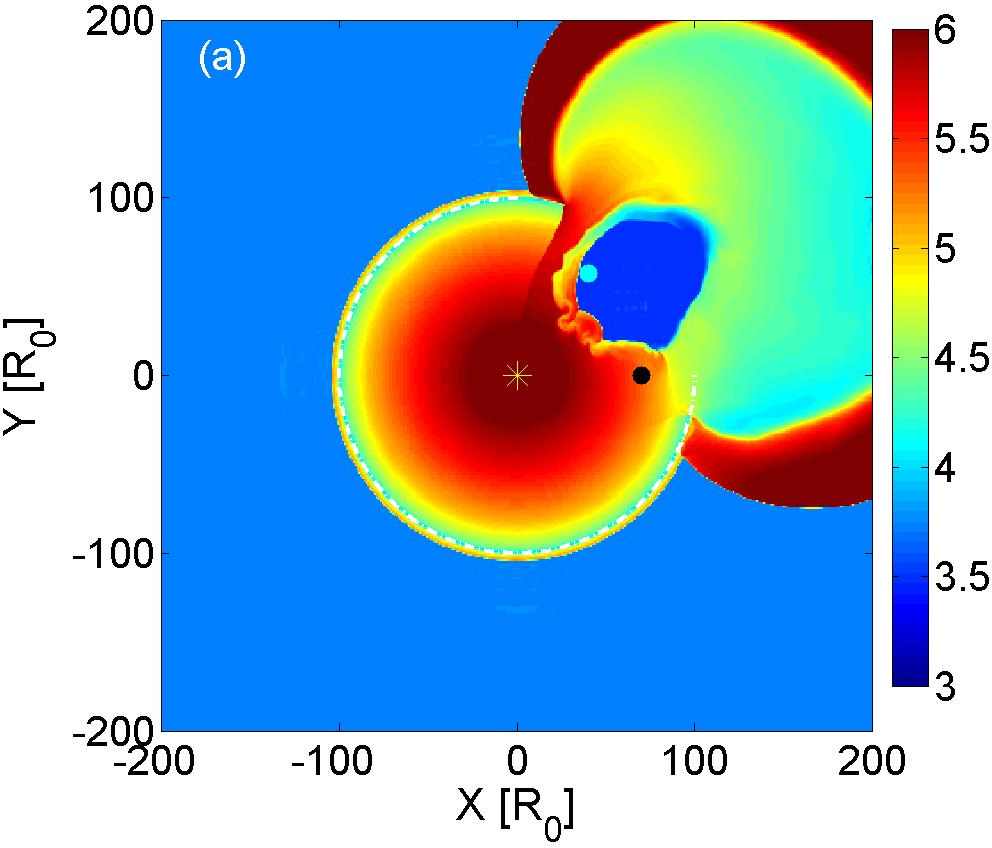}&
\includegraphics[width=6.5cm]{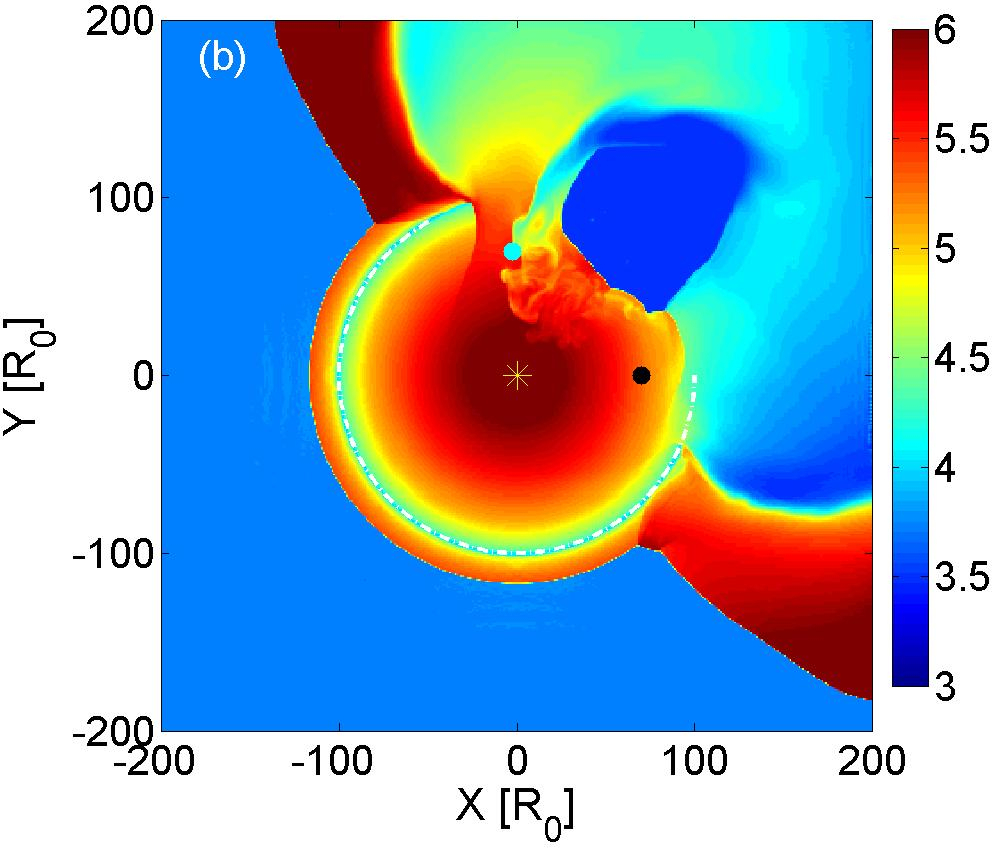}\\
\includegraphics[width=6.5cm]{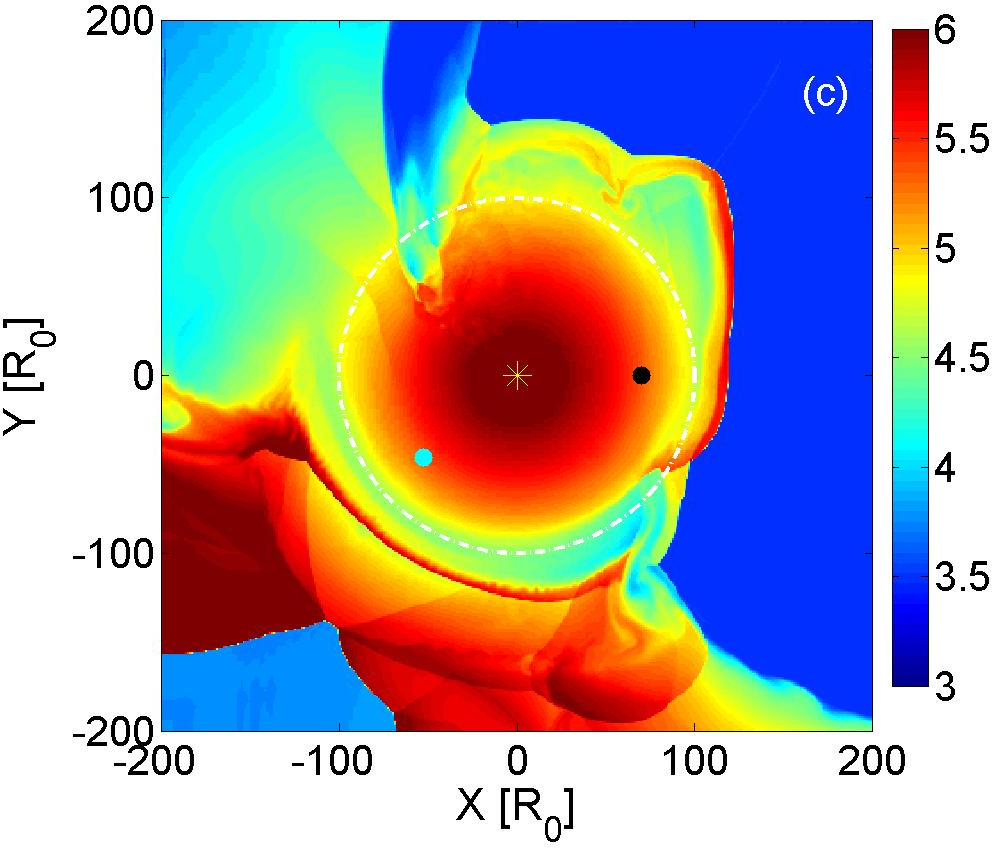}&
\includegraphics[width=6.5cm]{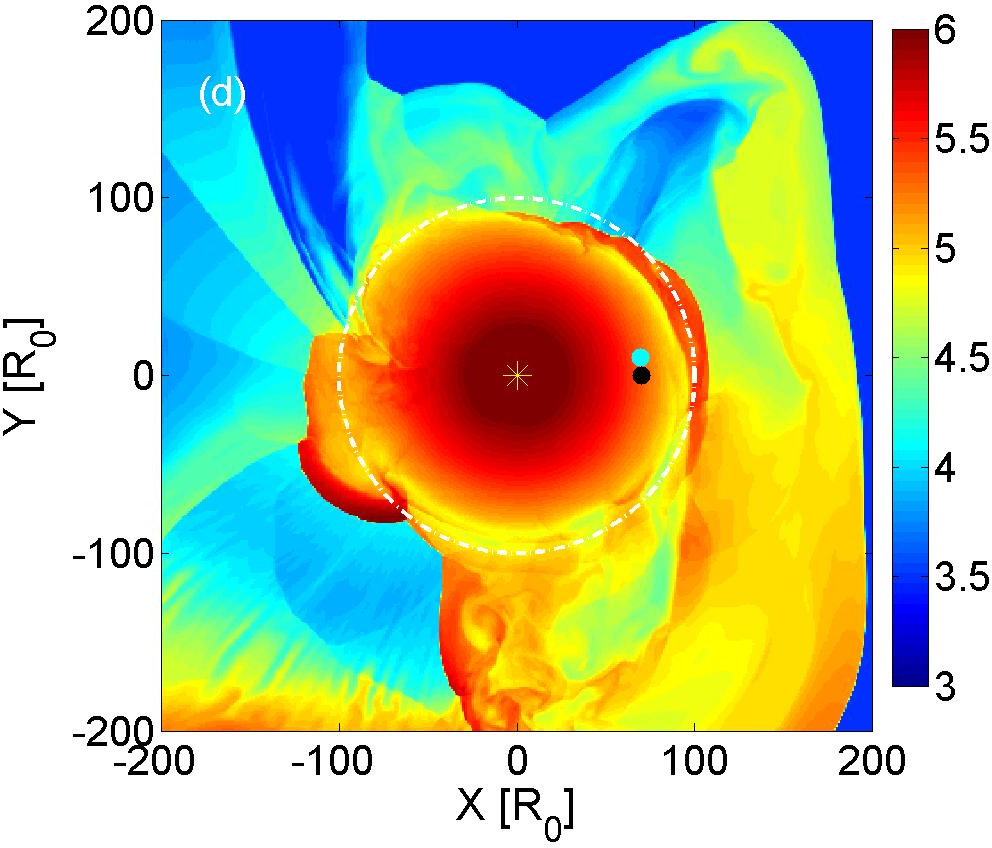}\\
\includegraphics[width=6.5cm]{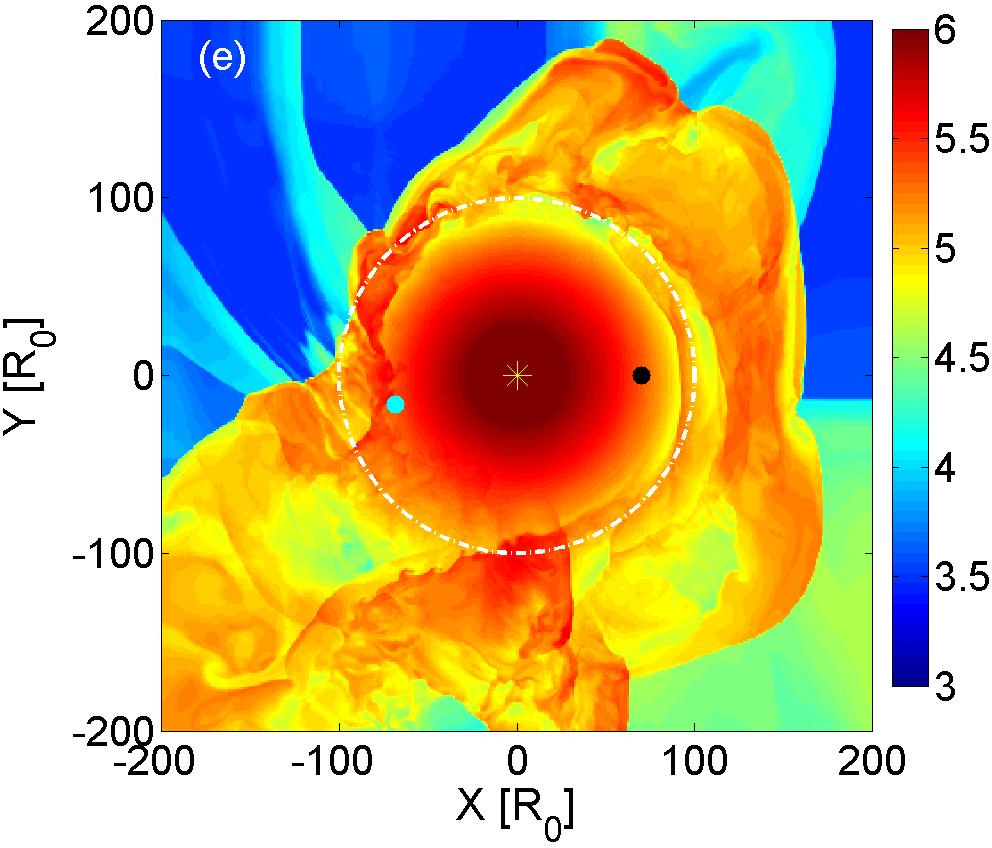}&
\includegraphics[width=6.5cm]{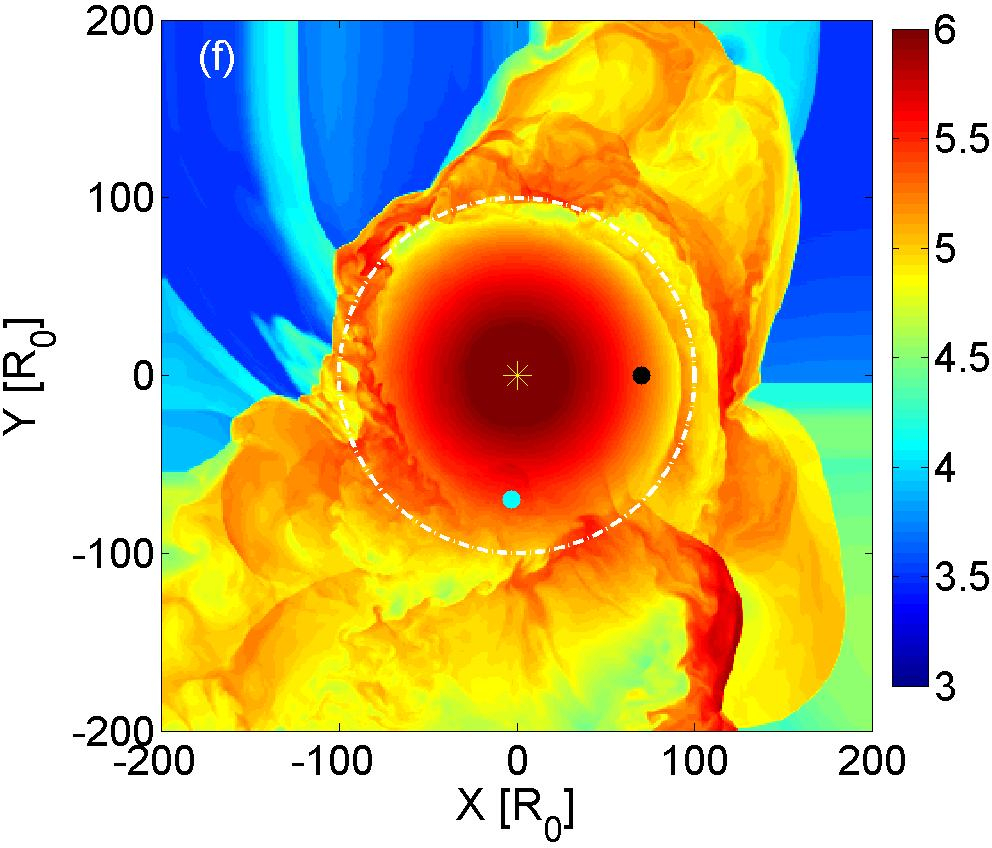}
\end{tabular}
\caption{Like Fig. \ref{fig:XYslices} but for the temperature in the orbital plane and at times of (a) $t=6$, (b) $t=9$, (c) $t=22$, (d) $t=36$, (e) $t=55$, and (f) $t=62 \days$.}
  \label{fig:XYtemperature}
    \end{figure}

We first present the outflow that results from the injection of mass and energy of the fiducial (low energy) run into the AGB envelope.
The mass and energy injection starts at $t=0$ at the location of the merger product at that time $(x,y,z)_{\rm mer-0}=(70 R_\odot,0,0)$, which we mark with a black dot in the figures.
The injection process lasts for a time of $T_{\rm mer} = 8.8$~hours, which is about 1 per cent of the orbital period.
In the first three figures (igs. \ref{fig:XYslices}-\ref{fig:XYtemperature}) we present some flow properties in the orbital plane $z=0$.
The merger product moves counterclockwise around the centre of the giant star,
and its location at each time is marked with a cyan dot.
% FFFFFFFFFFFFFFFFFFFFFFFFFFFFFFFFFFFFFFFFFFFFFFFFFF
% \includegraphics
% FFFFFFFFFFFFFFFFFFFFFFFFFFFFFFFFFFFFFFFFFFFFFFFFFF
% FFFFFFFFFFFFFFFFFFFFFFFFFFFFFFFFFFFFFFFFFFFFFFFFFF
% FFFFFFFFFFFFFFFFFFFFFFFFFFFFFFFFFFFFFFFFFFFFFFFFFF
% FFFFFFFFFFFFFFFFFFFFFFFFFFFFFFFFFFFFFFFFFFFFFFFFFF
% FFFFFFFFFFFFFFFFFFFFFFFFFFFFFFFFFFFFFFFFFFFFFFFFFF

From Figs. \ref{fig:XYslices}-\ref{fig:XYtemperature} we notice the following flow properties. The injection of the energy leads to a shock wave that propagates out, i.e., we basically set an explosion in the envelope.
The initial phase that lasts for about two days consists of pure expansion of the exploding sphere of mass and energy.
The injected mass suffers an adiabatic cooling.
This is seen as a low temperature region (blue) in
Fig. \ref{fig:XYtemperature} in the two upper panels.

The high temperature regions in the outer envelope are post-shock regions. Seen in  Fig. \ref{fig:XYtemperature}, a shock is propagating in the outer parts of the envelope around the centre.
When it closes on itself on the other side of the star it forms a gas prominence and high pressure region, as seen on the lower left corner of the two lower panels of Fig. \ref{fig:XYPressure}. The high temperature region can lead to a transient event, as we discuss later.

The blast wave that is formed by the injection of energy expands to all directions. About half of the injected material and the envelope mass that is directly pushed by the energy injection, move first toward the dense parts of the AGB envelope. The kinetic energy of this gas is dissipated in the envelope and does not eject mass to infinity. It rather drives shock waves that travel into and around the star and converge on its opposite side, as mentioned above.
The low density gas from near the merger site pushes onto the denser gas toward the centre and accelerates it inward, leading to the development of Rayleigh-Taylor (RT) instabilities. These are clearly seen in panels (b) and (c) of Fig. \ref{fig:XYslices}.

\begin{figure}[]
\centering
\begin{tabular}{cc}
\includegraphics[width=6.5cm]{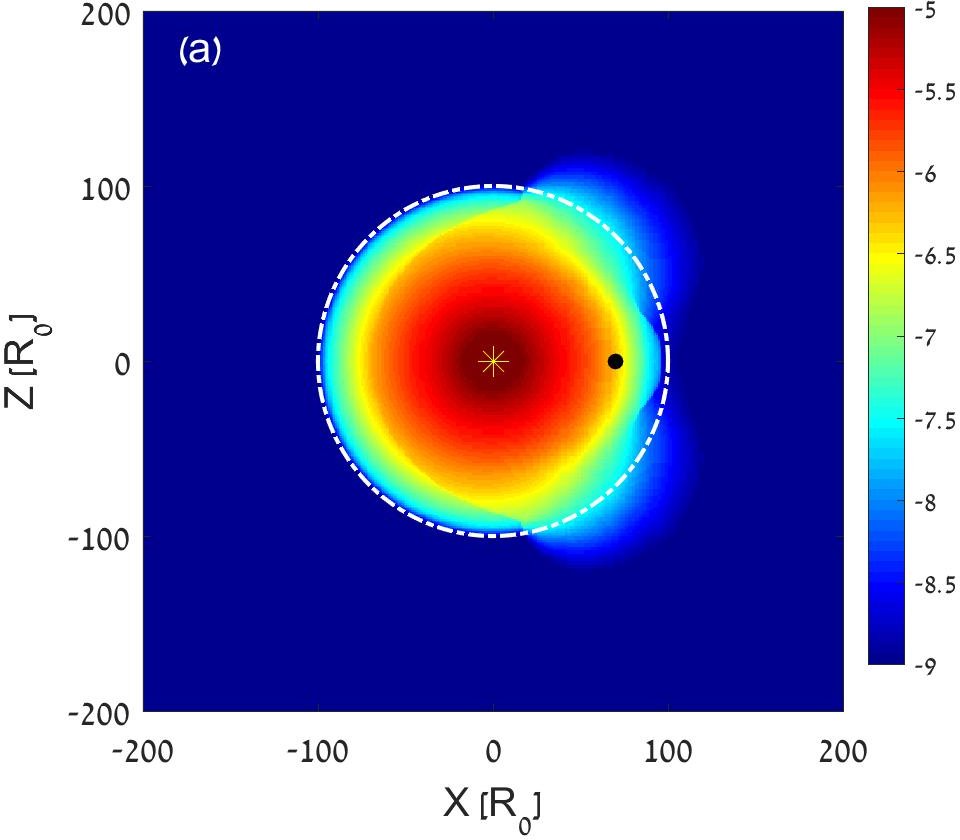}&
\includegraphics[width=6.5cm]{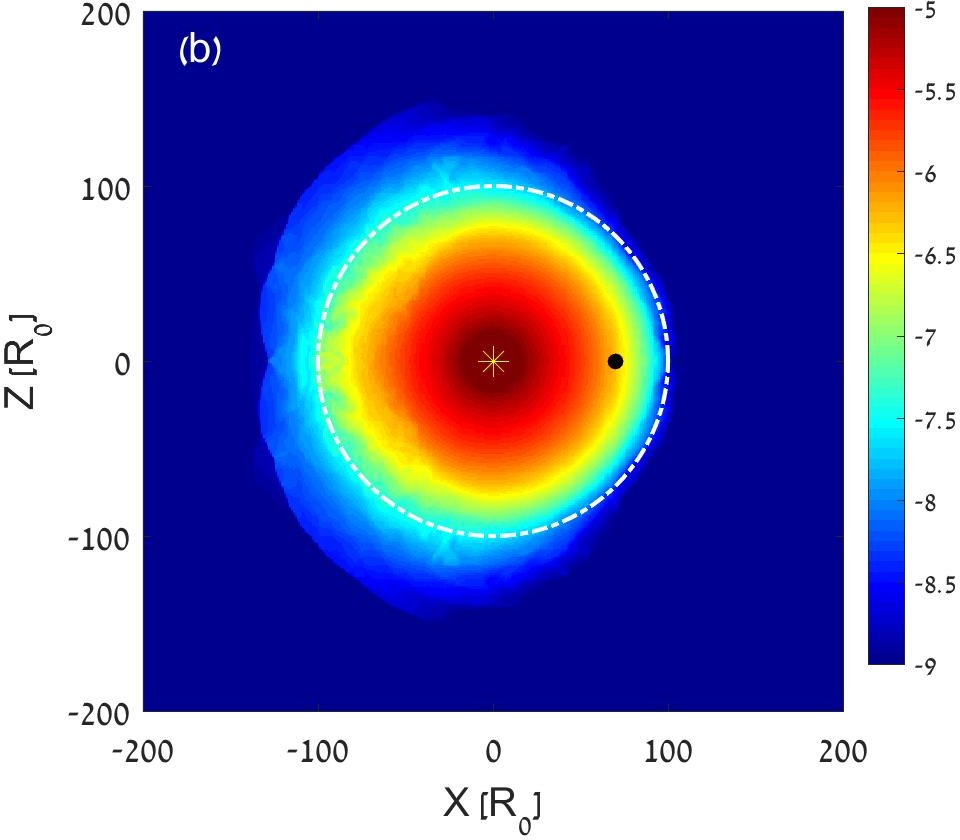}
\end{tabular}
\caption{Density in the meridional plane $y=0$ at
  (a) $t=13 \days$ and  (b) $t=44 \days$.
  The flow is symmetric about the equatorial plane, but highly asymmetric with respect to energy injection location (black dot), i.e., there is a large left-right asymmetry in the figure. }
    \label{fig:XZslices}
    \end{figure}

In Fig. \ref{fig:XZslices} we present the density in the meridional plane $y=0$, i.e., a plane perpendicular to the equatorial plane that goes through the centre of the AGB star. As expected, the flow is symmetric on the two sides of the equatorial plane. We can see the disturbance circling the star on both sides, i.e., above and below the equatorial plane.
% FFFFFFFFFFFFFFFFFFFFFFFFFFFFFFFFFFFFFFFFFFFFFFFFFF
% FFFFFFFFFFFFFFFFFFFFFFFFFFFFFFFFFFFFFFFFFFFFFFFFFF

In this first study where we ignore the gravity of the merger remnant we see only the influence of the injected energy and mass. We can see that the expanding gas lags behind the Keplerian motion of the merger product (the cyan dot on the figures).

By the end of this run, about $0.03\,M_\sun$ of gas had flown out of the numerical grid with a positive energy.
More material can escape the gravitational potential barrier with the aid of radiation pressure on dust that is expected to be formed in the cooling gas. As we do not include radiation pressure on the lifted gas, we somewhat underestimate the unbound mass.

\begin{figure}[]
\centering
\begin{tabular}{cc}
\includegraphics[width=6.5cm]{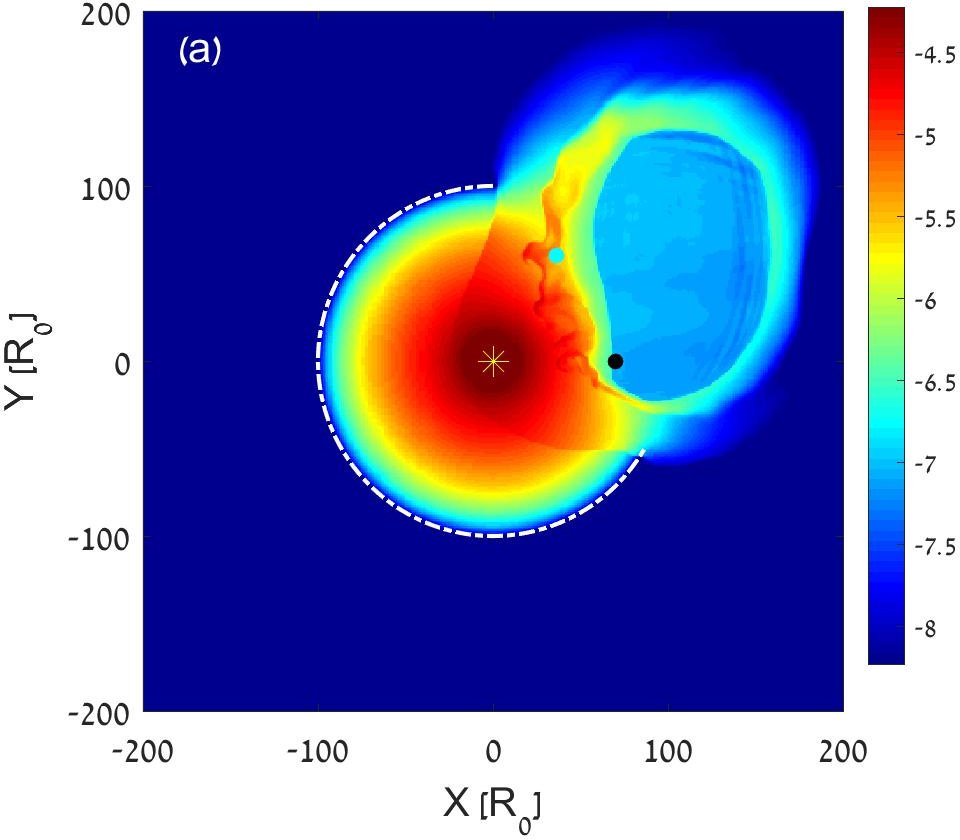}&
\includegraphics[width=6.5cm]{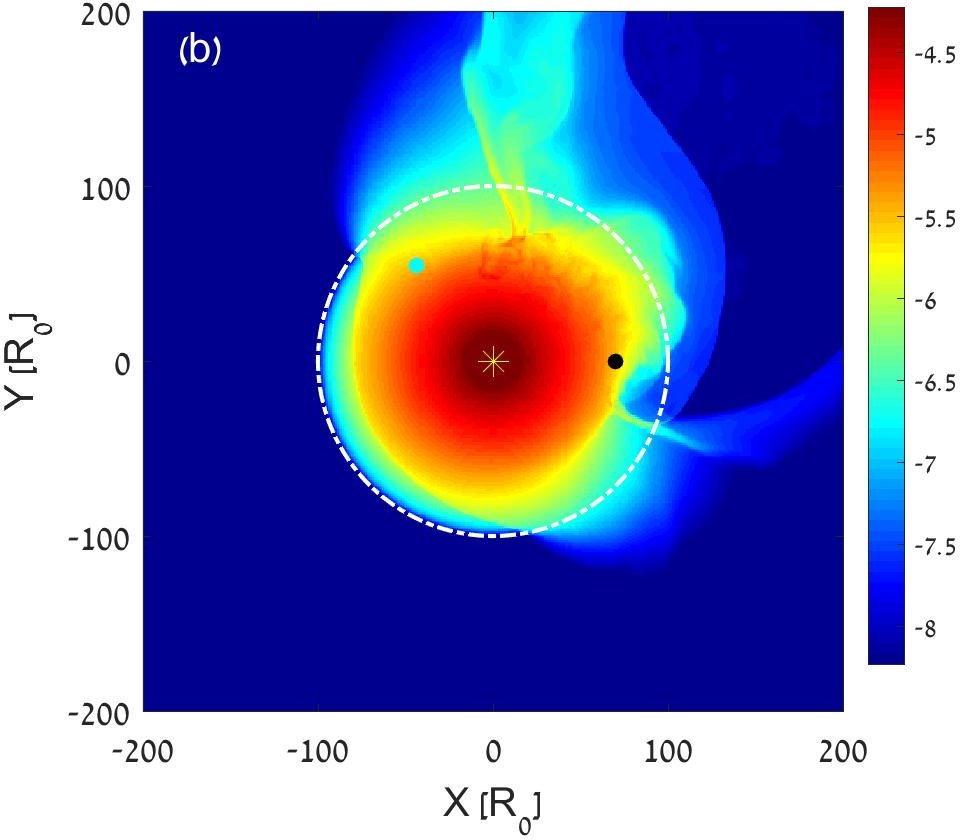}
\end{tabular}
\caption{Like Fig. \ref{fig:XYslices} but for the run with the higher injected energy of $E_{\rm mer} = 2.5 \times 10^{46} \erg$, and only at two times of  (a) $t=6 \days$ and  (b) $t=13 \days$. These are to be compared with panels b-d of Fig. \ref{fig:XYslices}}
  \label{fig:HighEnergy}
    \end{figure}

The injected energy of $E_{\rm mer} = 5 \times 10^{45} \erg$ can be accounted for even by the merger of two brown dwarfs. The merger of two low mass main sequence stars, of masses $M_1 \approx M_2 \simeq 0.1-0.2 M_\sun$ will release much larger amount of energy.
We simulated one such case with $E_{\rm mer} = 2.5 \times 10^{46} \erg$. The density in the equatorial plane at two times is presented in Fig. \ref{fig:HighEnergy}. Comparing to panels b-d in Fig. \ref{fig:XYslices}, we can immediately see that the AGB star suffers a much significant distortion. Due to numerical limitations we could not continue this run. It will be repeated in a new set of simulations that will include the self gravity of the envelope and the gravity of the merger product.
% FFFFFFFFFFFFFFFFFFFFFFFFFFFFFFFFFFFFFFFFFFFFFFFFFF
% FFFFFFFFFFFFFFFFFFFFFFFFFFFFFFFFFFFFFFFFFFFFFFFFFF

\begin{figure}[]
\centering
\begin{tabular}{cc}
%\vspace*{-0.5cm}
%\hskip -0.5 cm
\includegraphics[width=7.0cm]{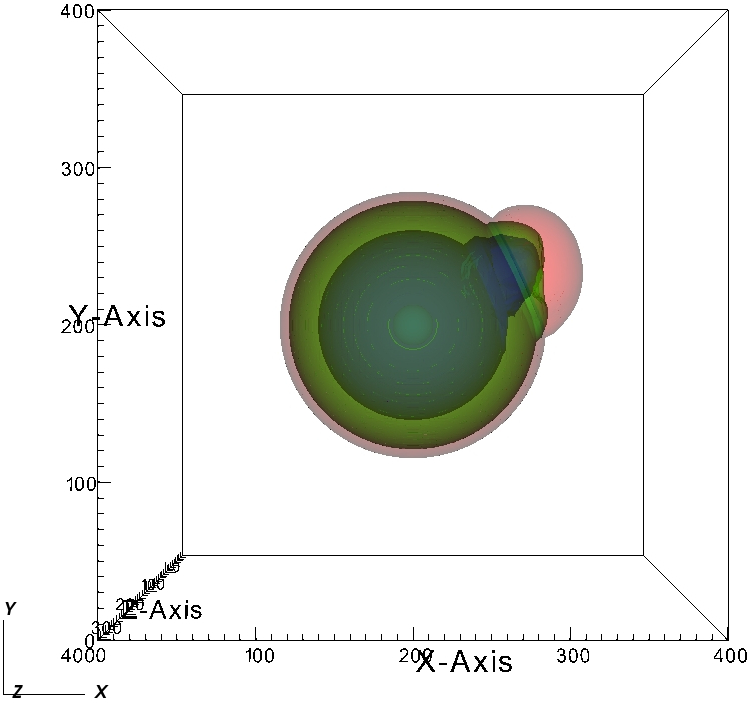}
%\hskip -1.5 cm
\includegraphics[width=7.0cm]{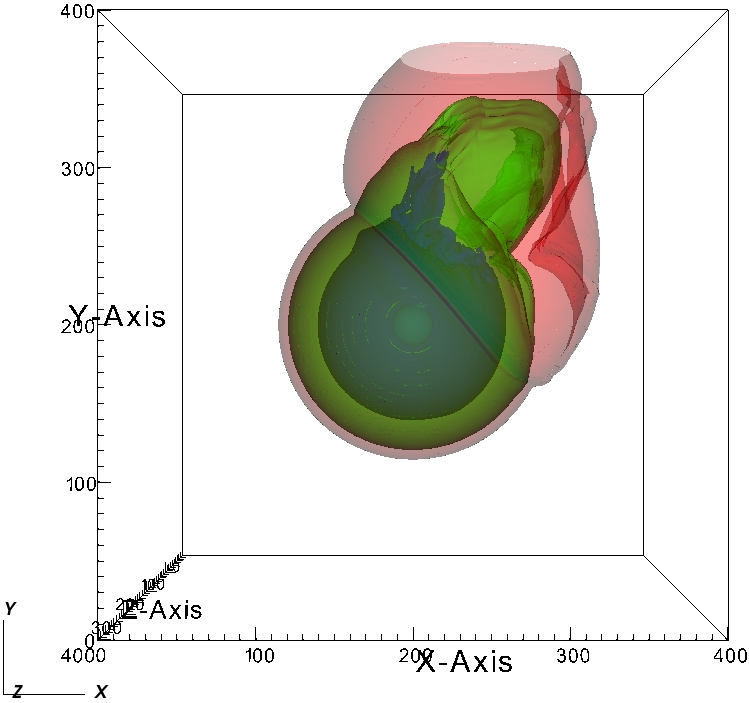} \\
%\vspace*{-0.5cm}
%\hskip -0.5 cm
\includegraphics[width=7.0cm]{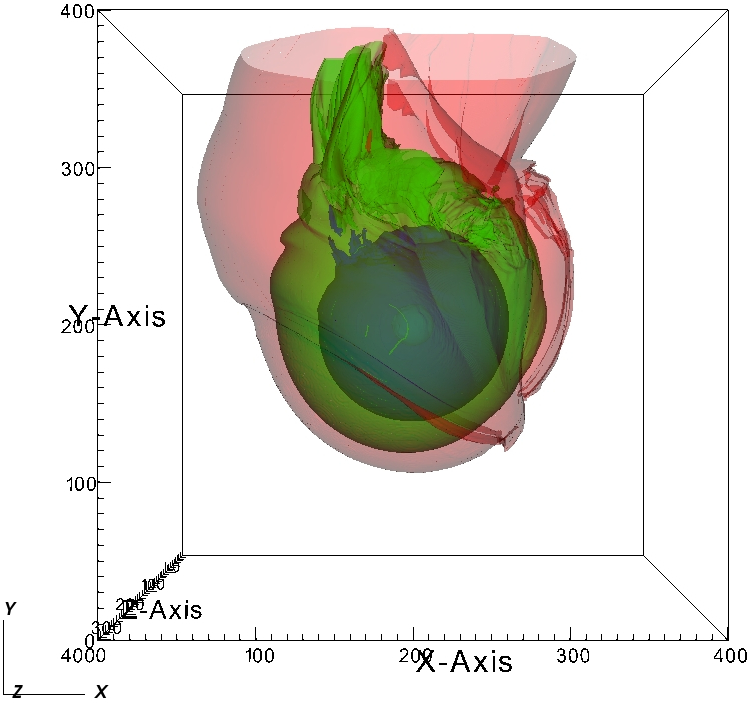}
%\hskip -1.5 cm
\includegraphics[width=7.0cm]{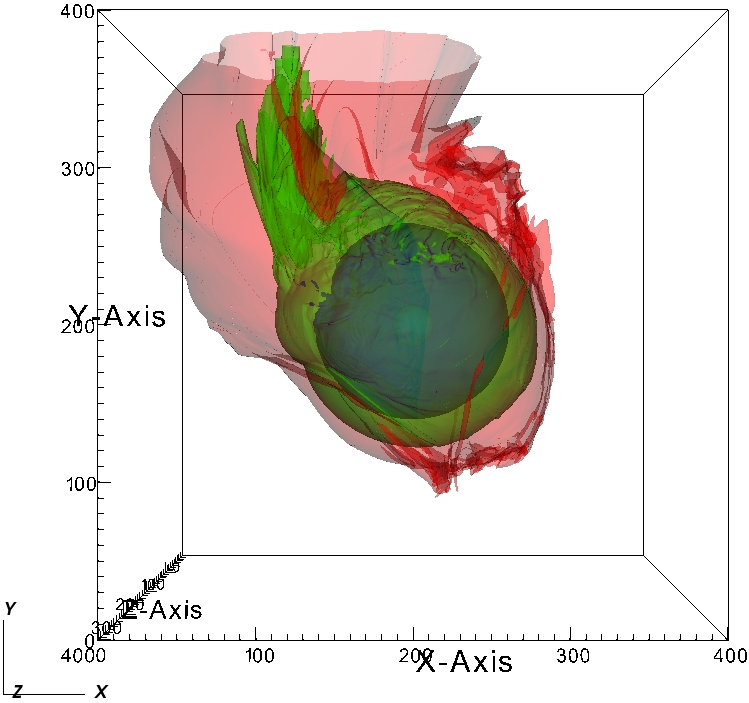} \\
%\vspace*{-0.5cm}
%\hskip -0.5 cm
\includegraphics[width=7.0cm]{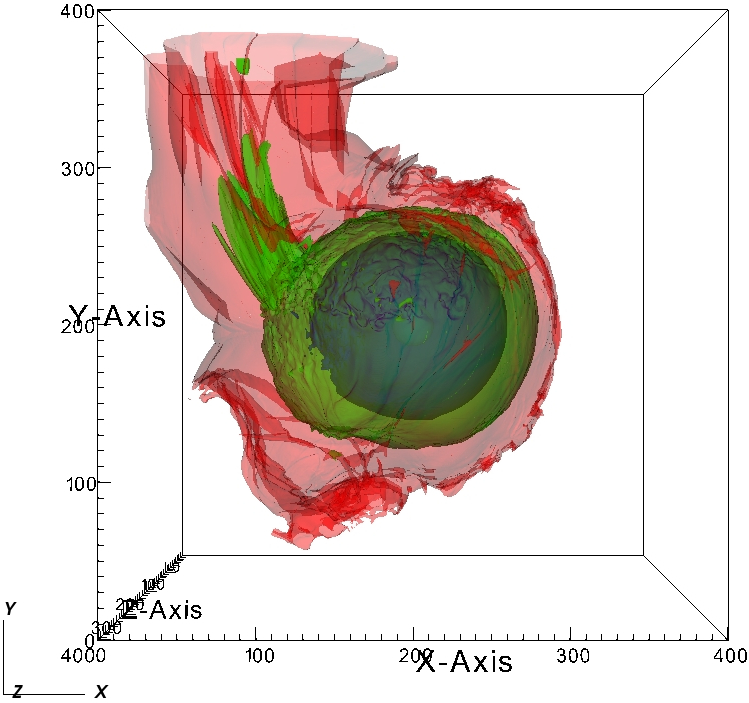}
%\hskip -1.5 cm
\includegraphics[width=7.0cm]{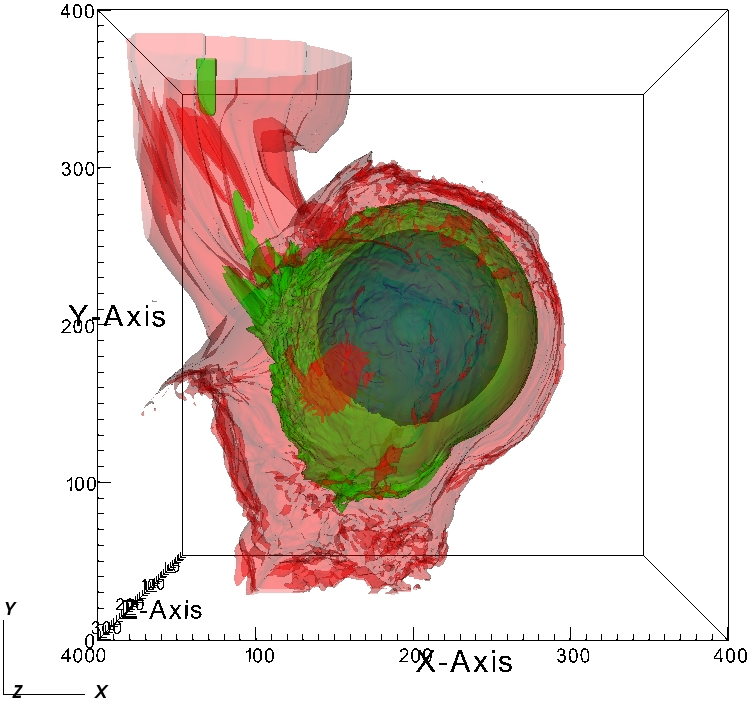}
\end{tabular}
\caption{ Three dimensional density structure for the low energy case at times of (from upper left to lower right) (a) $t=4$, (b) $t=11$, (c) $t=22$, (d) $t=33$ (e) $t=44$, and (f) $t=55 \days$.
Shown are equi-density surfaces, with colour-coding of red $6\times 10^{-9}\g \cm^{-3}$, green $1.2\times 10^{-7}\g \cm^{-3}$,
blue $2.8\times 10^{-6}\g \cm^{-3}$, and pale blue $6\times 10^{-5}\g \cm^{-3}$.
}
  \label{fig:3D_density}
    \end{figure}

The full distortion of the AGB envelope is seen the best in 3D images.
In Fig. \ref{fig:3D_density} we present equi-density surfaces at six times. The formation of a messy circumstellar matter is clearly seen, mainly by following the red colour that depicts a density surface of $\rho= 6 \times 10^{-9}\g \cm^{-3}$.
% FFFFFFFFFFFFFFFFFFFFFFFFFFFFFFFFFFFFFFFFFFFFFFFFFF
% FFFFFFFFFFFFFFFFFFFFFFFFFFFFFFFFFFFFFFFFFFFFFFFFFF

\begin{figure}[h]
\centering
\begin{tabular}{c}
\includegraphics[width=12cm]{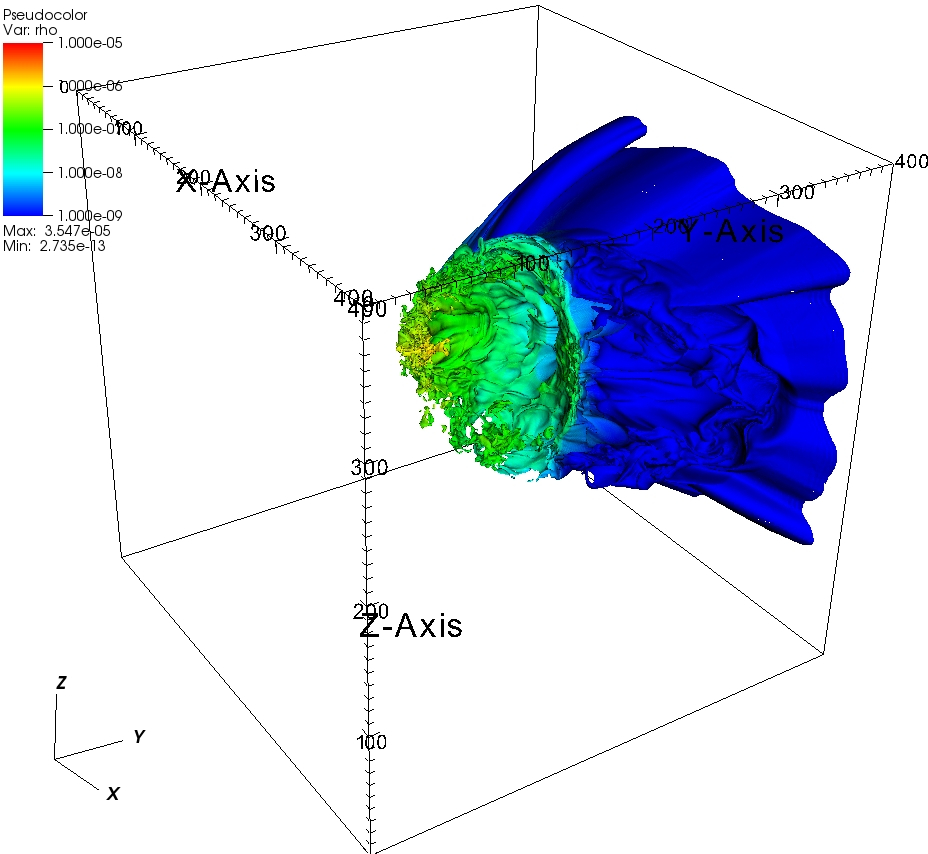}
\end{tabular}
\caption{The boundary between the injected material and the gas of the giant star at $t = 37 \days$. Presented is the surface where half of the mass in each grid point originated from the injected merger mass. Namely, the surface of a tracer value of $\xi=0.5$. The colours represent the total density on that surface, from $\rho=10^{-9} \g \cm^{-1}$ blue, through $\rho=10^{-7} \g \cm^{-1}$ (green)  to $\rho=10^{-6} \g \cm^{-1}$ (yellow). }
\label{fig:tracer}
\end{figure}

We end by presenting the 3D structure of the gas that was injected in the merger process. To follow this gas we use a `tracer', which is a non-physical variable moving with specifically designated gas, the injected mass in the present case.
The tracer's value at each point indicates the mass fraction of the injected gas at that point.
Thus, regions occupied by the gas ejected from the merger process are marked by a tracer value of $\xi=1$, and regions where the ejected gas has not reached are marked by $\xi=0$.
Figure \ref{fig:tracer} depicts the surface corresponding to a tracer value of $\xi=0.5$ at $t = 37 \days$.
This surface delineates the boundary between the gas ejected in the merger process and its surroundings.
The colour of each point on the surface indicates the local density. The expanding gas ejected from the merger process can not penetrate the dense layers of the AGB envelope, and it expands radially outwards in a direction opposite the centre.
At a late time the gas can be seen to be concentrated, very crudely, on a surface of a pyramid.
There is a small azimuthal velocity component due to the orbital motion during the merger process.
% FFFFFFFFFFFFFFFFFFFFFFFFFFFFFFFFFFFFFFFFFFFFFFFFFF
% FFFFFFFFFFFFFFFFFFFFFFFFFFFFFFFFFFFFFFFFFFFFFFFFFF

%==========================================================
\section{DISCUSSION AND SUMMARY}
\label{sec:summary}
% ==========================================================

Many rare transient events, some that include the explosion of stars and some that include the complete destruction of stars by another object, have been discovered in recent years. Many more are expected with coming telescopes and surveys. The goal of this study is to present one such very rare event that, nonetheless, might one day be detected as an ILOT. As well, this process can lead to the formation of a very asymmetrical (messy) circumstellar matter, e.g., a messy PN. The process is that of a tight  binary system that enters the envelope of a giant star. Because of the friction inside the envelope, the two stars of the tight binary system merge and release energy and mass inside the envelope. The tight binary system can be composed of any two non-giant stars, e.g., neutron stars, white dwarfs, main sequence stars, horizontal branch stars, and/or brown dwarfs, and any combination of these. The secondary star might even be a massive planet.

 In the present study, still preliminary, we simply injected energy inside the envelope during a very short time (about 9 hours) relative to the orbital period (about 36 days). We included the gravitational field of the giant star, but we kept it constant at its initial value. To isolate the influence of the inserted energy, and due to limited computer resources, we neglect in this first study the following processes. We did not consider the influence of the tight binary system on the structure of the envelope prior to the merger. The envelope is expected to rotate and to have an oblate structure. We started at $t=0$ with a spherical giant envelope. We did not include the gravity of the merger product and the changing gravity of the deformed envelope (self-gravity). The merger processes might lead to the formation of an accretion disk around the surviving star. This disk might launch jets. We did not study this process here. We plan future simulations where we will include these effects. Another effect that we did not include, and which is very complicated to include, is radiative transfer.

We present the evolution of the flow for the low-energy (fiducial) simulation in Figs. \ref{fig:XYslices} - \ref{fig:XZslices}, and in the 3D images in Figs. \ref{fig:3D_density} and \ref{fig:tracer}.
We note that even the merger of two brown dwarfs can release the injected energy of $E_{\rm mer} = 5 \times 10^{45} \erg$ in this case.
We summarize the main results of this simulation as follows.

\textit{(1) An ILOT.}
A shock propagates on the outskirts of the giant envelope. This is best seen by the deep red colour in Fig. \ref{fig:XYtemperature}. In addition to adiabatic cooling, the hot gas is expected to cool by radiation. This will lead to an outburst in the visible that will evolve to red as dust is formed in the ejecta. This ILOT will last for several weeks. The exact light curve depends on the energy injected and how deep inside the envelope it is injected. This is not studied in the present paper.

\textit{ (2) A very asymmetrical mass ejection.}
The flow that results from the merger process does not have a symmetry around an axis that goes through the core and the location of the merger process because the energy is being injected with the velocity of the tight binary system around the giant star. As can be see in the different figures, a messy outflow is formed. This will lead to the formation of a messy nebula, and hence this is one of the triple stellar evolutionary routes that lead to the formation of messy PNe \citep{Soker2016triple, BearSoker2017}.
In the present simulations the flow possesses a symmetry about the orbital plane. However, in reality, the orbital plane of the tight binary system does not need to be in the orbital plane of the triple stellar system. In that case there will be a departure from a plane symmetry as well.

\textit{(3) Ejected mass. }
The shock that propagates through the giant envelope causes prominences and ejection of mass for about several weeks. Part of the mass leaves the grid with a positive energy, and hence it will escape the system. In the simulation that we conducted, a relatively small amount of energy was deposited within the giant envelope, and hence only about $0.03 M_\sun$ has a positive energy and leaves the system. Although not much mass, it is enough to leave an imprint on the descendant nebula.

We run one case with five times larger injected energy.
In Fig. \ref{fig:HighEnergy} we present the first 13 days of the simulation with a larger injected energy of $E_{\rm mer} = 2.5 \times 10^{46} \erg$. As expected, the envelope is much more distorted and the outflow from the star is messy already at this early time. Due to numerical limitations we could not follow this highly distorted envelope. We will study such violent outbursts in a future paper where we will modify the numerical code to include the self-gravity of the highly distorted envelope.

A general and a wider short summary of our study can be phrased as follows. Rare triple stellar evolution routes, such as the one studied here, extend the domain of the binary interaction to account for different types of astrophysical objects, like ILOTs and shaping of circumstellar nebulae, and strengthen the binary interaction model.

\section*{Acknowledgments}
{{{ We thank an anonymous referee for useful comments. }}}
We thank Efrat Sabach for her help with the AGB model, and Michael Refaelovich for his assistance.
This research was supported by the Prof. A. Pazy Research Foundation.
N.S. is supported by the Charles Wolfson Academic Chair.

% %%%%%%%%%%%%  Refrences %%%%%%%%%%%%%%%%%%%%%%%%%%%%%%%%%%%%%%%%%%%%%%%%%%%%%%%%%%%%%%%%%%%%%%%%%%%%%%%%%%%%%

\label{lastpage}


\begin{thebibliography}{}\addcontentsline{toc}{section}{References}


\bibitem[Akashi \& Soker(2008)]{AkashiSoker2008} Akashi, M., \& Soker, N.\ 2008, \mnras, 391, 1063

\bibitem[Akashi \& Soker(2017)]{AkashiSoker2017} Akashi, M., \& Soker, N.\ 2017, arXiv:1701.05460

\bibitem[Ali et al.(2016)]{Alietal2016} Ali, A., Dopita, M.~A., Basurah, H.~M., Amer, M.~A., Alsulami, R., \& Alruhaili, A.\ 2016, \mnras, 462, 1393

\bibitem[Akras et al.(2015)]{Akrasetal2015} Akras, S., Boumis, P.,
Meaburn, J., Alikakos, J., Lopez, J. A., Goncalves, D.~R.\ 2015,
\mnras, 452, 2911

\bibitem[Akras et al.(2016)]{Akrasetal2016} Akras, S., Clyne, N., Boumis, P., Monteiro, H., Goncalves, D.~R., Redman, M.~P., \& Williams, S.\ 2016, \mnras, 457, 3409

\bibitem[Aller et al.(2015a)]{Alleretal2015a} Aller, A., Miranda,
L.~F., Olgu{\'{\i}}n, L., Vazquez, R., Guillen, P.~F., Oreiro, R.,
Ulla, A., \& Solano, E.\ 2015a, \mnras, 446, 317

\bibitem[Aller et al.(2015b)]{Alleretal2015b} Aller, A., Montesinos,
B., Miranda, L.~F., Solano, E., \& Ulla, A.\ 2015b, \mnras, 448,
2822

\bibitem[Balick et al.(2013)]{Balicketal2013} Balick, B., Huarte-Espinosa, M., Frank, A., Gomez, T., Alcolea, J., Corradi, R.~L.~M., \& Vinkovic, D.\ 2013, \apj, 772, 20

\bibitem[Bear \& Soker(2017)]{BearSoker2017} Bear, E., \& Soker, N.\ 2017, \apjl, 837, L10

\bibitem[Boffin(2015)]{Boffin2015} Boffin, H.\ 2015, 19th European
Workshop on White Dwarfs, 493, 527

\bibitem[Boffin et al.(2012)]{Boffinetal2012} Boffin, H.~M.~J., Miszalski, B., Rauch, T.,  Jones, D., Corradi, R.~L.~M., Napiwotzki, R., Day-Jones, A.~C., \& K\"oppen, J.\ 2012, Science, 338, 773

\bibitem[Bond(2000)]{Bond2000} Bond, H.~E.\ 2000, Asymmetrical
Planetary Nebulae II: From Origins to Microstructures, 199, 115

\bibitem[Bond et al.(2016)]{Bondetal2016} Bond, H.~E., Ciardullo, R., Esplin, T.~L., Hawley, S.~A., Liebert, J., \& Munari, U.\ 2016, \apj, 826, 139

\bibitem[Bond et al.(1978)]{Bondetal1978} Bond, H.~E., Liller, W., \& Mannery, E.~J.\ 1978, \apj, 223, 252

\bibitem[Bond \& Livio(1990)]{BondLivio1990} Bond, H.~E., \& Livio, M.\
1990, \apj, 355, 568

\bibitem[Bond et al.(2002)]{Bondetal2002} Bond, H.~E., O'Brien,
M.~S., Sion, E.~M., Mullan, D.~J., Exter, K., Pollacco, D.~L., \&
Webbink, R.~F.\ 2002, Exotic Stars as Challenges to Evolution,
279, 239

\bibitem[Chen et al.(2017)]{Chenetal2017} Chen, Z., Frank, A., Blackman, E.~G., Nordhaus, J., \& Carroll-Nellenback, J.\ 2017, arXiv:1702.06160

\bibitem[Chen et al.(2016)]{Chenetal2016} Chen, Z., Nordhaus, J., Frank, A., Blackman, E.~G., \& Balick, B.\ 2016, \mnras, 460, 4182

\bibitem[Chiotellis et al.(2016)]{Chiotellisetal2016} Chiotellis, A., Boumis, P., Nanouris, N., Meaburn, J., \& Dimitriadis, G.\ 2016, \mnras, 457, 9

\bibitem[Corradi et al.(2015)]{Corradietal2015} Corradi, R.~L.~M.,
Garc{\'{\i}}a-Rojas, J., Jones, D., \& Rodr{\'{\i}}guez-Gil, P.\
2015, \apj, 803, 99

\bibitem[Danehkar et al.(2013)]{Danehkaretal2013} Danehkar, A., Parker, Q.~A., \& Ercolano, B.\ 2013, \mnras, 434, 1513

\bibitem[Decin et al.(2015)]{Decinetal2015} Decin, L., Richards, A.~M.~S.,
Neufeld, D., Steffen, W., Melnick, G., \& Lombaert, R.\ 2015,
\aap, 574, A5

\bibitem[De Marco(2015)]{DeMarco2015} De Marco, O.\ 2015, in Physics of Evolved Stars - A conference dedicated to the memory of Olivier Chesneau, Eds. E. Lagadec, F. Millour and T. Lanz, EAS Publications Series, 71, 357

\bibitem[De Marco et al.(2015)]{DeMarcoetal2015} De Marco, O., Long,
J., Jacoby, G.~H., Hillwig, T., Kronberger, M., Howell, S.~B.,
Reindl, N., Margheim, S.\ 2015, \mnras, 448, 3587

\bibitem[De Marco \& Soker(2011)]{DeMarcoSoker2011} De Marco, O., \& Soker, N.\
2011, \pasp, 123, 402

\bibitem[Douchin et al.(2015)]{Douchinetal2015} Douchin, D., De Marco,
O., Frew, D.~J., Jacoby, G.~H., Jasniewicz, G., Fitzgerald, M.,
Passy, J-C., Harmer, D., Hillwig, T., \& Moe, M.\ 2015, \mnras,
448, 3132

\bibitem[Eggleton \& Verbunt(1986)]{EggletonVerbunt1986} Eggleton, P.~P., \& Verbunt, F.\ 1986, \mnras, 220, 13P

\bibitem[Exter et al.(2010)]{Exteretal2010} Exter, K., Bond, H.~E., Stassun, K.~G., Smalley, B., Maxted, P.~F.~L., \& Pollacco, D. L.\ 2010, \aj, 140, 1414

\bibitem[Fabian \& Hansen(1979)]{FabianHansen1979} Fabian, A.~C., \& Hansen, C.~J.\ 1979, \mnras, 187, 283

\bibitem[Fang et al.(2015)]{Fangetal2015} Fang, X., Guerrero, M.~A.,
Miranda, L.~F., Riera, A., Velazquez, P.~F., Raga, A. C.\ 2015,
\mnras, 452, 2445

\bibitem[Garc{\'{\i}}a-Rojas et al.(2016)]{GarciaRojasetal2016} Garc{\'{\i}}a-Rojas, J., Corradi, R.~L.~M., Monteiro, H., Jones, D., Rodriguez-Gil, P., \& Cabrera-Lavers, A.\ 2016, \apjl, 824, L27

\bibitem[Garc{\'{\i}}a-Segura et al.(2014)]{GarciaSeguraetal2014}
Garc{\'{\i}}a-Segura, G., Villaver, E., Langer, N., Yoon, S.-C.,
\& Manchado, A.\ 2014, \apj, 783, 74

\bibitem[Gorlova et al.(2015)]{Gorlovaetal2015} Gorlova, N., Van Winckel, H., Ikonnikova, N.~P., Burlak, M.~A., Komissarova, G.~V., Jorissen, A., Gielen, C., Debosscher, J., \& Degroote, P.\ 2015, \mnras, 451, 2462

\bibitem[Han et al.(1995)]{Hanetal1995} Han, Z., Podsiadlowski, P., \& Eggleton, P.~P.\ 1995, \mnras, 272, 800

\bibitem[Hillwig et al.(2016a)]{Hillwigetal2016a} Hillwig, T.~C., Bond, H.~E., Frew, D.~J., Schaub, S.~C., \& Bodman, E.~H.~L.\ 2016a, \aj, 152, 34

\bibitem[Hillwig et al.(2015)]{Hillwigetal2015} Hillwig, T.~C., Frew, D.~J., Louie, M.,  De Marco, O., Bond, H.~E., Jones, D., Schaub, S.~C.\ 2015, \aj, 150, 30

\bibitem[Hillwig et al.(2016b)]{Hillwigetal2016b} Hillwig, T., Jones, D., De Marco, O., Bond, H., Margheim, S., \& Frew, D.\ 2016b,
\apj, 832, 125

\bibitem[Huang et al.(2016)]{Huangetal2016} Huang, P.-S., Lee, C.-F., Moraghan, A., \& Smith, M.\ 2016, \apj, 820, 134

\bibitem[Huarte-Espinosa et al.(2012)]{HuarteEspinosaetal2012} Huarte-Espinosa, M., Frank, A., Balick, B., Blackman, E.~G., De Marco, O., Kastner, J.~H., \& Sahai, R. \ 2012, \mnras, 424, 2055
\bibitem[Humphreys et al.(1999)]{Humphreysetal1999} Humphreys, R.~M., Davidson, K., \& Smith, N.\ 1999, \pasp, 111, 1124

\bibitem[Iben \& Tutukov(1989)]{IbenTutukov1989} Iben, I., Jr., \& Tutukov, A.~V.\ 1989, Planetary Nebulae, 131, 505

\bibitem[Jones(2015)]{Jones2015} Jones, D.\ 2015, EAS Publications Series, 71, 113

\bibitem[Jones(2016)]{Jones2016} Jones, D.\ 2016, Journal of Physics Conference Series, 728, 032014

\bibitem[Jones \& Boffin(2017a)]{JonesBoffin2017} Jones, D., \& Boffin, H.~M.~J.\ 2017a, \mnras, 466, 2034

\bibitem[Jones \& Boffin(2017b)]{JonesBoffin2017b} Jones, D., \& Boffin, H.~M.~J.\ 2017b, Nature Astronomy 1, 0117

\bibitem[Jones et al.(2015)]{Jonesetal2015} Jones, D., Boffin, H.~M.~J., Rodr{\'{\i}}guez-Gil, P., Wesson, R., Corradi, R.~L.~M., Miszalski, B., \& Mohamed, S.\ 2015, \aap, 580, A19

\bibitem[Jones et al.(2017)]{Jonesetal2017} Jones, D., Van Winckel, H., Aller, A., Exter, K., \& De Marco, O.\ 2017, arXiv:1703.05096

\bibitem[Jones et al.(2016)]{Jonesetal2016} Jones, D., Wesson, R., Garc{\'{\i}}a-Rojas, J., Corradi, R.~L.~M., \& Boffin, H.~M.~J.\ 2016, \mnras, 455, 3263

\bibitem[Kashi \& Soker(2010)]{KashiSoker2010} Kashi, A., \& Soker, N.\ 2010, \apj, 723, 602

\bibitem[Kiminki et al.(2016)]{Kiminkietal2016} Kiminki, M.~M., Reiter, M., \& Smith, N.\ 2016, \mnras,

\bibitem[Livio \& Shaviv(1975)]{LivioShaviv1975} Livio, M., \& Shaviv, G.\ 1975, \nat, 258, 308

\bibitem[Madappatt et al.(2016)]{Madappattetal2016} Madappatt, N., De Marco, O., \& Villaver, E.\ 2016, \mnras,

\bibitem[Manick et al.(2015)]{Manicketal2015} Manick, R., Miszalski,
B., \& McBride, V.\ 2015, \mnras, 448, 1789

\bibitem[Mart{\'{\i}}nez Gonz{\'a}lez et al.(2015)]{Martinezetal2015} Mart{\'{\i}}nez Gonz{\'a}lez,
M.~J., Asensio Ramos, A., Manso Sainz, R., Corradi, R.~L.~M., \&
Leone, F.\ 2015, \aap, 574, A16

\bibitem[Michaely \& Perets(2014)]{MichaelyPerets2014} Michaely, E., \& Perets, H.~B.\ 2014, \apj, 794, 122

\bibitem[Mignone et al.(2007)]{Mignone2007} Mignone, A., Bodo, G., Massaglia, S., et al.\ 2007, \apjs, 170, 228

\bibitem[Miszalski et al.(2013)]{Miszalskietal2013} Miszalski, B., Boffin, H.~M.~J., \& Corradi, R.~L.~M.\ 2013, \mnras, 428, L39

\bibitem[Miszalski et al.(2015)]{Miszalskietal2015} Miszalski, B., Manick, R., \& McBride, V.\ 2015, in Physics of Evolved Stars - A conference dedicated to the memory of Olivier Chesneau, Eds. E. Lagadec, F. Millour and T. Lanz, EAS Publications Series, 71, 117 (arXiv:1507.07707)

\bibitem[Mo{\v c}nik et al.(2015)]{Mocniketal2015} Mo{\v c}nik, T.,
Lloyd, M., Pollacco, D., \& Street, R.~A.\ 2015, \mnras, 451, 870

\bibitem[Montez et al.(2015)]{Montezetal2015} Montez, R., Jr.,
Kastner, J.~H., Balick, B., et al.\ 2015, \apj, 800, 8

\bibitem[Morris(1981)]{Morris1981} Morris, M.\ 1981, \apj, 249, 572

\bibitem[Morris(1987)]{Morris1987} Morris, M.\ 1987, \pasp, 99, 1115

\bibitem[Morris \& Podsiadlowski(2009)]{MorrisPodsiadlowski2009} Morris, T., \& Podsiadlowski, P.\ 2009, \mnras, 399, 515

\bibitem[Nordhaus \& Blackman(2006)]{NordhausBlackman2006} Nordhaus, J., \&
Blackman, E.~G.\ 2006, \mnras, 370, 2004

\bibitem[Paxton et al.(2011)] {Paxtonetal2011} Paxton, B., Bildsten, L., Dotter, A., et al. 2011, \apjs, 192, 3

\bibitem[Paxton et al.(2013)] {Paxtonetal2013} Paxton, B., Cantiello, M., Arras, P., et al. 2013, \apjs, 208, 4

\bibitem[Paxton et al.(2015)]{Paxtonetal2015} Paxton, B., Marchant, P., Schwab, J., et al.\ 2015, \apjs, 220, 15

\bibitem[Paczynski(1985)]{Paczynski1985} Paczynski, B.\ 1985, Cataclysmic Variables and Low-Mass X-ray Binaries, 113, 1

\bibitem[Portegies Zwart \& van den Heuvel(2016)]{Portegies2016} Portegies Zwart, S.~F., \& van den Heuvel, E.~P.~J.\ 2016, \mnras, 456, 3401

\bibitem[Rechy-Garc{\'{\i}}a et al.(2017)]{RechyGarciaetal2017}  Rechy-Garc{\'{\i}}a, J., Vel{\'a}zquez, P.~F., Pe{\~n}a, M., \& Raga, A.~C.\ 2017, \mnras, 464, 2318

\bibitem[Sahai et al.(2016)]{Sahaietal2016} Sahai, R., Scibelli, S., \& Morris, M.~R.\ 2016, \apj, 827, 92

\bibitem[Sahai \& Trauger(1998)]{SahaiTrauger1998} Sahai, R., \& Trauger, J.~T.\ 1998, \aj, 116, 1357

\bibitem[Soker(1990)]{Soker1990AJ} Soker, N.\ 1990, \aj, 99, 1869

\bibitem[Soker(1994)]{Soker1994} Soker, N.\ 1994, \mnras, 270, 774

\bibitem[Soker(2004)]{Soker2004} Soker, N.\ 2004, \mnras, 350, 1366

\bibitem[Soker(2016)]{Soker2016triple} Soker, N.\ 2016, \mnras, 455, 1584

\bibitem[Soker \& Hadar(2002)]{SokerHadar2002} Soker, N., \& Hadar, R.\ 2002, \mnras, 331, 731

\bibitem[Soker \& Harpaz(1992)]{SokerHarpaz1992} Soker, N., \& Harpaz, A.\
1992, \pasp, 104, 923

\bibitem[Soker et al.(1992)]{Sokeretal1992} Soker, N., Zucker, D.~B., \& Balick, B.\ 1992, \aj, 104, 2151

\bibitem[Tocknell et al.(2014)]{Tocknelletal2014} Tocknell, J., De Marco, O., \& Wardle, M.\ 2014, \mnras, 439, 2014

\bibitem[Zijlstra(2015)]{Zijlstra2015} Zijlstra, A.~A.\ 2015, \rmxaa, 51, 221



\end{thebibliography}
\end{document}